\documentclass{article}

\usepackage{graphicx}%
\usepackage{multirow}%
\usepackage{amsmath,amssymb,amsfonts}%
\usepackage{amsthm}%
\usepackage{mathrsfs}%
\usepackage[title]{appendix}%
\usepackage{xcolor}%
\usepackage{textcomp}%
\usepackage{manyfoot}%
\usepackage{booktabs}%
\usepackage{algorithm}%
\usepackage{algorithmicx}%
\usepackage{algpseudocode}%
\usepackage{hyperref}
\usepackage{listings}%
\usepackage[numbers,sort&compress]{natbib}
\usepackage[english]{babel}
\usepackage{multirow}
\usepackage{rotating}
\usepackage{mathtools}
\usepackage{adjustbox}
\usepackage{enumitem}
\setlist{nosep}
\usepackage{physics}
\usepackage{gensymb}
\usepackage{lmodern}
\usepackage{tabularx}
\usepackage{fix-cm}
\usepackage{mathrsfs}
\usepackage{anyfontsize}
\usepackage{authblk}
\usepackage[textfont=up]{caption}
\DeclareCaptionFormat{plain}{\normalfont#1#2#3\par}
\newcommand{\orgname}[1]{#1}
\newcommand{\orgaddress}[1]{#1}
\newcommand{\street}[1]{#1}
\newcommand{\city}[1]{#1}
\newcommand{\state}[1]{#1}
\newcommand{\postcode}[1]{#1}
\newcommand{\country}[1]{#1}
\newcommand{\department}[1]{\textit{#1}}
\newcommand{\keywords}[1]{\paragraph*{Keywords:} #1}
\date{}

\begin{document}

\title{Quantum Algorithm for Protein Structure Prediction Using the Face-Centered Cubic Lattice}


\newcommand*\firstauth[1][\value{footnote}]{\footnotemark[#1]\textsuperscript{,}}

\author[1]{Rui-Hao Li\thanks{These authors contributed equally to this work and are designated as co-first authors.}\textsuperscript{,}}
\author[2]{Hakan Doga\firstauth}
\author[1]{Bryan Raubenolt\firstauth}
\author[2]{Sarah Mostame}
\author[3]{Nicholas DiSanto}
\author[1]{Fabio Cumbo}
\author[1]{Jayadev Joshi}
\author[1]{Hanna Linn}
\author[1]{Maeve Gaffney}
\author[1]{Alexander Holden}
\author[1,4]{Vinooth Kulkarni}
\author[4]{Vipin Chaudhary}
\author[1]{Kenneth M. Merz Jr}
\author[2]{Abdullah Ash Saki}
\author[5]{Tomas Radivoyevitch}
\author[3]{Frank DiFilippo}
\author[6]{Jun Qin}
\author[2]{Omar Shehab}
\author[1,7]{Daniel Blankenberg\thanks{To whom correspondence should be addressed: blanked2@ccf.org}\textsuperscript{,}}

\affil[1]{\department{Center for Computational Life Sciences}, \orgname{Cleveland Clinic}, \orgaddress{\street{9500 Euclid Ave}, \city{Cleveland}, \state{OH}, \postcode{44195}, \country{USA}}}
\affil[2]{IBM Quantum, IBM Thomas J Watson Research Center, Yorktown Heights, NY 10598, USA}
\affil[3]{\department{Department of Nuclear Medicine}, \orgname{Cleveland Clinic}, \orgaddress{\street{9500 Euclid Ave}, \city{Cleveland}, \state{OH}, \postcode{44195}, \country{USA}}}
\affil[4]{\department{Department of Computer and Data Sciences}, \orgname{Case Western Reserve University}, \orgaddress{ \city{Cleveland}, \state{OH}, \postcode{44106}, \country{USA}}}
\affil[5]{\department{Department of Quantitative Health Sciences}, \orgname{Cleveland Clinic}, \orgaddress{\street{9500 Euclid Ave}, \city{Cleveland}, \state{OH}, \postcode{44195}, \country{USA}}}
\affil[6]{\department{Department of Cardiovascular \& Metabolic Sciences}, \orgname{Cleveland Clinic}, \orgaddress{\street{9500 Euclid Ave}, \city{Cleveland}, \state{OH}, \postcode{44195}, \country{USA}}}
\affil[7]{\department{Department of Molecular Medicine}, \orgname{Cleveland Clinic Lerner College of Medicine, Case Western Reserve University}, \orgaddress{\street{9501 Euclid Ave}, \city{Cleveland}, \state{OH}, \postcode{44195}, \country{USA}}}

\maketitle

\begin{abstract}
     In this work, we present the first implementation of the face-centered cubic (FCC) lattice model for protein structure prediction with a quantum algorithm. Our motivation to encode the FCC lattice stems from our observation that the FCC lattice is more capable in terms of modeling realistic secondary structures in proteins compared to other lattices, as demonstrated using root mean square deviation (RMSD). We utilize two quantum methods to solve this problem: a polynomial fitting approach (PolyFit) and the Variational Quantum Eigensolver with constraints (VQEC) based on the Lagrangian duality principle. Both methods are successfully deployed on Eagle R3 (\texttt{ibm\_cleveland}) and Heron R2 (\texttt{ibm\_kingston}) quantum computers, where we are able to recover ground state configurations for the 6-amino acid sequence \texttt{KLVFFA} under noise. A comparative analysis of the outcomes generated by the two QPUs reveals a significant enhancement (reaching nearly a two-fold improvement for PolyFit and a three-fold improvement for VQEC) in the prediction and sampling of the optimal solution (ground state conformations) on the newer Heron R2 architecture, highlighting the impact of quantum hardware advancements for this application.

\end{abstract}



\section{Introduction}
Protein structure prediction (PSP) is a field of interest to many disciplines of life sciences and medicine.  
Biological activities are often a manifestation of interactions between proteins, functions of which are directly related to the three-dimensional (3D) structures they adopt in specific physiological environments. 
Understanding fundamental mechanisms of health and disease relies on knowing, or predicting, these structures.  
Computational models offer an approach to solving this problem.  
Indeed, the field of template-based modeling (TM), where machine learning algorithms are trained on available experimental structures, has provided some of the most successful PSP methods.  
The inherent reliance on these experimental structures, which are produced at a much slower rate than the identification of the corresponding primary amino acid sequences (through genomic sequencing), will continue to be a limitation for these methods.  
Free modeling (FM) methods, in which most of the effort is placed on simulating the physics responsible for yielding these folded structures in nature (such as molecular dynamics simulations and other \textit{ab initio} PSP methods) continue to be explored as viable alternatives, as they arguably require less of a direct dependence on a database of homologous proteins.  
Given the computational complexity and scalability limitations of classical FM methods, a growing field has emerged wherein FM approaches are being developed using quantum algorithms (herein referred to as QPSP methods). 

\begin{figure}[t]
    \centering
    \includegraphics[width=14.5cm]{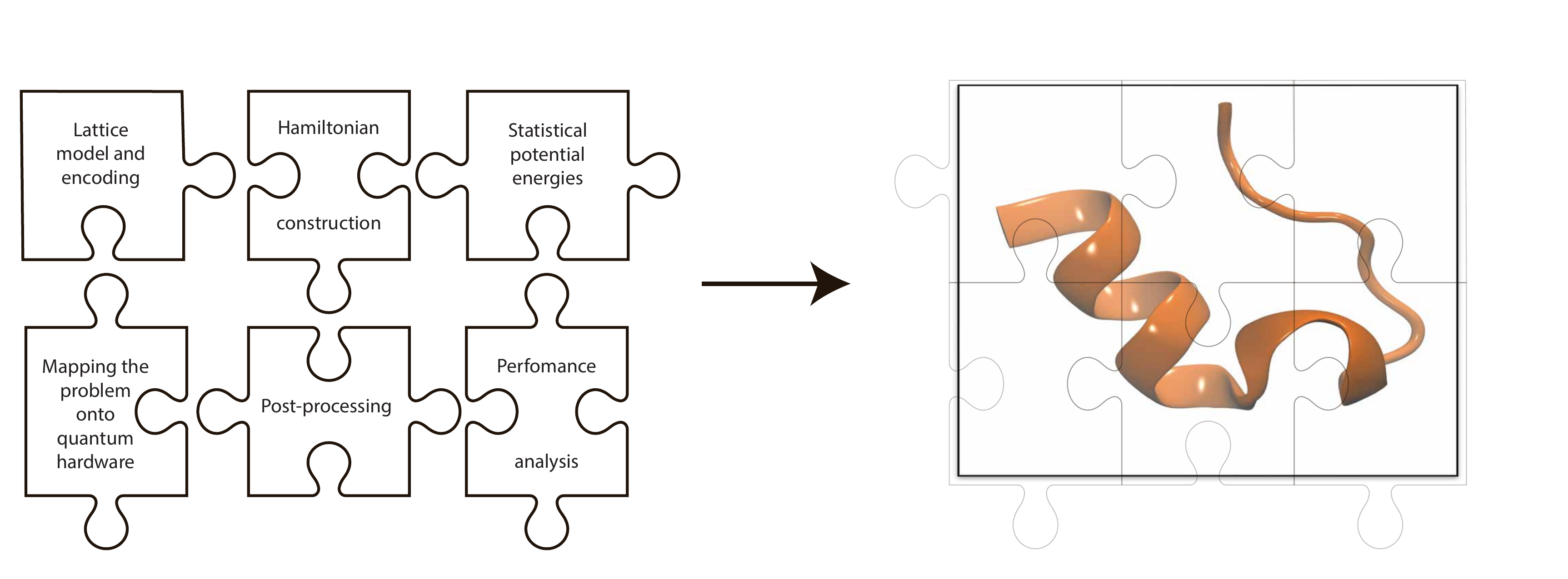}
    \caption{\small From the selection of an optimal lattice model to final performance analysis, protein structure prediction with quantum algorithms (QPSP) is a complex task that requires a careful analysis of each of these individual pieces. Depicted here is the Trp-cage miniprotein (PDB: 2JOF \cite{Barua2008}).}
    \label{puzzle_workflow}
\end{figure}

In most quantum FM methods, conformational spaces are generally discretized onto coarse-grained lattices, and residues are reduced to single spheres centered at each amino acid's alpha carbon (C$\alpha$). 
Knowledge-based scoring functions, also known as statistical potentials, are used to differentiate conformers.  
While this inherently implies these are lower resolution models (compared to many explicit all-atom classical FM methods), the obvious benefit is the resulting quantum resource efficiency, which is always a considerable factor especially with pre-fault tolerant quantum hardware.  
With this in mind, to successfully predict the structure of a protein using a quantum computer, several pieces need to align well. 
For example, there is a delicate dependence of these models on the construction of the Hamiltonian terms, including the statistical potentials used, and the choice and encoding of a lattice.  
Optimal solutions are fully determined by the problem Hamiltonian formulations.  
Aside from this, there are other important components unique to quantum computing, such as the choice of ansatz in variational quantum algorithms (VQAs) for efficient sampling and the classical optimizer chosen.
Finding a delicate balance between quantum resources, e.g., qubit count, Hamiltonian locality, etc., and accuracy of the prediction, is a major challenge in the QPSP space.
To this end, one of the most resource-efficient QPSP methods is formulated on a tetrahedral lattice~\cite{robertResourceefficientQuantumAlgorithm2021}.
However, as we will show, this lattice is not very flexible; it is limited in the number of turn angles it can generate, and the number of directions in which the protein chain can grow at each lattice point.
Thus, it is not suitable for predicting secondary structures such as $\alpha$-helices, where the turn angles are notably different from the tetrahedral lattice's 109.5\degree.
Moreover, there is considerable post-processing needed to generate all-atom biological structures that can be compared to experimental structures to assess model performance.  
Prior work that utilized QPSP has placed less emphasis on the latter, which is a differentiating factor in our work.  Part of our ongoing effort involves optimizing the use of bioinformatics approaches to automate this post-processing, and will be presented in future works.
Fig.~\ref{puzzle_workflow} illustrates how these pieces interconnect to produce a fully realized protein structure prediction. 

In this work, we consider a more granular lattice, the face-centered cubic (FCC) lattice, which has not been previously implemented in QPSP.
Due to its higher degree of freedom and non-bipartite nature, the encoding of the FCC lattice is more complex than that of the tetrahedral lattice and requires more qubits.
We present a quantum encoding of the FCC lattice, focusing on two new methods of constructing the problem Hamiltonian that avoid using slack variables, which are often used to enforce inequality constraints but can lead to an increase in the number of qubits and terms in the problem Hamiltonian.
We further analyze the trade-off between the number of qubits and number of terms in the problem Hamiltonian for these methods. 
Our implementation of the FCC lattice is motivated by our observation that this lattice, compared to earlier coarse-grained lattices, is capable of capturing more realistic biological structures such as secondary structures. 
Finally, we present our quantum hardware experiment results executed on \texttt{ibm\_cleveland} and \texttt{ibm\_kingston}, highlighting the improvements on the newer Heron processor.

\section{A brief review of coarse-grained lattice models in QPSP}

The field of QPSP is a relatively young field compared to its classical counterparts. 
Various coarse-grained lattice models in 2D and 3D for structure prediction such as the cubic and tetrahedral lattices, have been proposed and implemented for quantum computing.
The QPSP problem is formulated via discretization of the configuration (or conformational) space, in which the structure prediction is treated as an optimization problem. 
Finding the optimal structure is then reduced to finding the ground state of a Hamiltonian that encodes the energy of the protein structure.
The Hamiltonian is constructed from a set of terms that represent the interactions between the residues in the protein chain, as well as various constraints that ensure the validity of the structure and nearest neighbor pairs.
Earlier works explored different lattice models, and various encoding methods to achieve more realistic representations of the biological protein structures. 
In tandem with this, we observe a shift towards the use of 3D lattices instead of 2D, and more efficient encoding of these lattice models, such as turn-based encoding instead of coordinate-based, reducing the number of qubits required and controlling the scaling properties of the problem Hamiltonian. 
However, in reality, protein structures are not restricted by these discrete, pre-defined lattice coordinates, which limits the ability for many of these models to approximate native state structures in practice. 
As a result, the accuracy of lattice-based protein structure predictions is ultimately limited by the resolution of the lattice, with lattice models permitting higher degrees of freedom generally yielding more realistic fits~\cite{Godzik1993}.
We provide a summary table of the prior works in this field in App.~\ref{app:prior_work}, including the lattice models used, the encoding methods, and other relevant details (see Table~\ref{lattice_models}).

The quantum algorithm developed by \citet{robertResourceefficientQuantumAlgorithm2021} for the tetrahedral lattice model is a significant improvement in terms of reducing the quantum resources needed to formulate the structure prediction problem, while maintaining a biologically faithful representation. 
Of all the prior works in this field, with the exception of \citet{num_13}, which employed a Lennard-Jones (LJ) cost function~\cite{Schwerdtfeger2024} on an all-heavy atom backbone model rather than only alpha carbons (C$\alpha$), the tetrahedral lattice was arguably the most realistic model given the permissible dihedral angles (180\degree, +60\degree, and -60\degree).  
Despite this resource efficiency, which makes it favorable for pre-fault tolerant quantum computers, the limitation also stems from the fact that the protein chain can grow into one of only 4 directions at each lattice point.  Moreover, only one angle can be formed between two consecutive turns, 109.5\degree. 
With this in mind, we performed extensive testing on this lattice in an effort to verify whether it remains suitable for the prediction of secondary structures such as $\alpha$-helices, and compared it with the best fits obtained using the FCC lattice (see Fig.~\ref{bestfits}).

\subsection{Why the FCC lattice?}

\begin{figure}[htbp]
    \centering
    \includegraphics[scale=0.65]{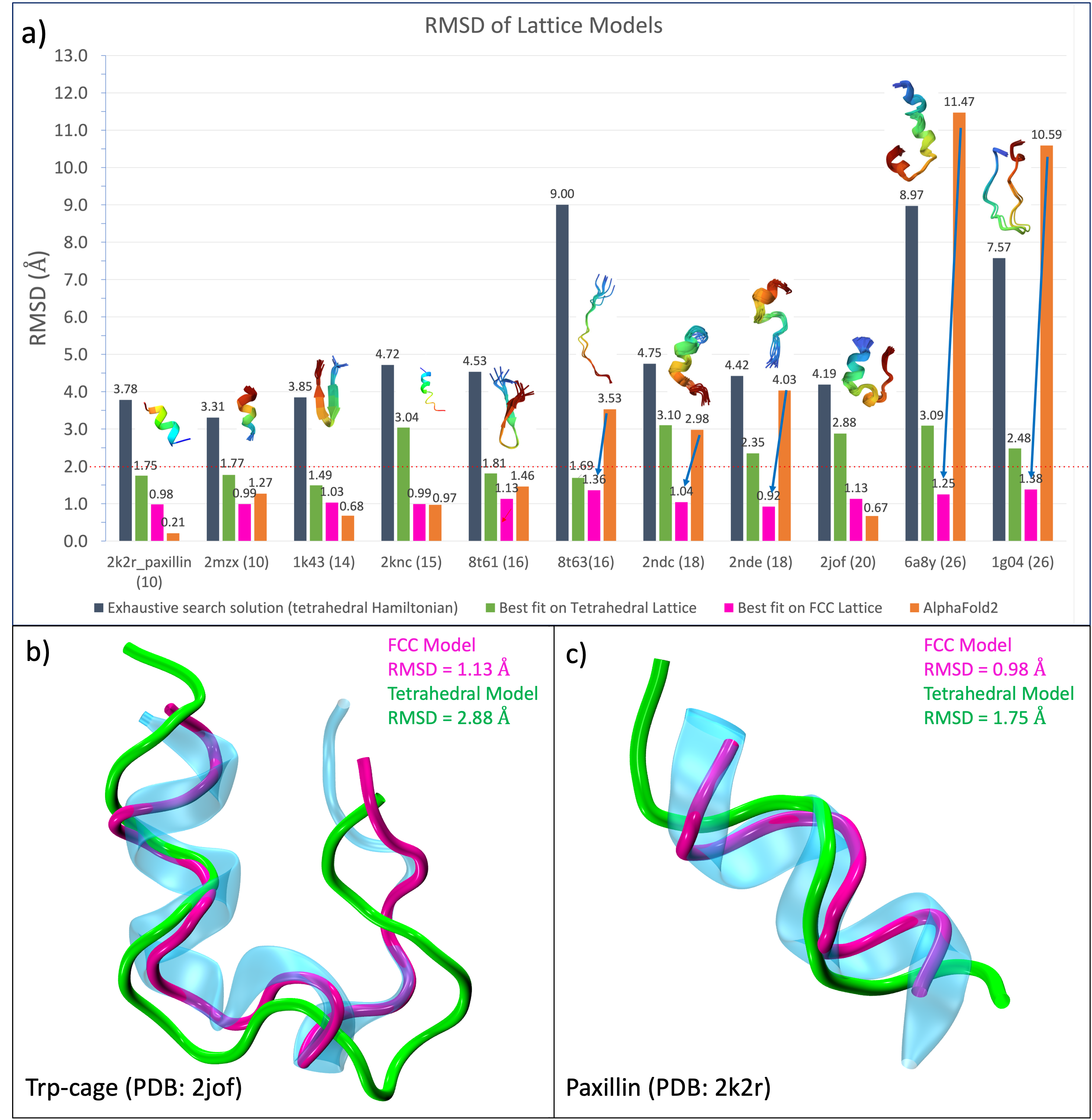}
    \caption{\small a) Compared are the original Hamiltonian's~\cite{robertResourceefficientQuantumAlgorithm2021} best solution using a tetrahedral lattice (grey), the best possible fit on a tetrahedral lattice (green), the best possible fit on an FCC lattice (magenta), and AlphaFold2 (orange). RMSDs were measured after alignment to known structures.  For the NMR structures, models were fit against the first frame of the ensemble. All known structures were obtained from the PDB\cite{pdb}, with the exception of a 15 AA integrin fragment (labeled as Jun Qin integrin) solved by one of team members. Visual examples of the differences in the models are provided for b) Trp-cage and c) paxillin. In both cases, the experimental structure is colored in transparent cyan, the tetrahedral lattice model in green, and the fcc lattice model in magenta.}
    \label{bestfits}
\end{figure}

While the C$\alpha$-only tetrahedral lattice has its benefits over the standard cubic lattices, the limited number of basis vectors results in discrepancies in modeling secondary structures, in particular $\alpha$-helices~\cite{Godzik1993}.  
Prior works have highlighted the greater accuracy when using lattices with higher degrees of freedom, and specifically the face-centered-cubic (FCC) lattice~\cite{Godzik1993,Park1995}.  
The FCC lattice provides higher resolution, allowing for 12 directions of growth and 4 angles between any two consecutive turns: 60\degree, 90\degree, 120\degree, and 180\degree. Moreover, the FCC lattice has been shown to result in significantly lower energy solutions \cite{clote2008protein} compared to other coarse-grained lattices, on the famous ``Harvard instances" \cite{yue1996folding} that contain 10 different benchmarking proteins. As a result, the novel quantum encoding of the FCC lattice that we present here can open up possibilities for more realistic modeling of the protein structures among other QPSP methods.

While the FCC lattice model is more capable in terms of capturing more realistic protein structures, the trade-off is the quantum resources needed to encode and solve this problem, in particular the number of qubits and the size of the problem Hamiltonian (see Sec.~\ref{fcc_section} for further details). Based on our prior resource estimates on the tetrahedral lattice~\cite{doga2024perspective}, and our estimates with FCC (see Fig.~\ref{fig:qubit_counts}), we carefully selected experimentally determined protein structures from the Protein Data Bank (PDB)~\cite{pdb} ranging in size between 10 and 26 amino acid residues, with enough structural diversity to include $\alpha$-helices, $\beta$-sheets, disordered loops, and combinations of these as well. 
Fig.~\ref{bestfits}a demonstrates our initial analysis on these structures.
The data here includes the top solutions on the tetrahedral lattice in gray (found by performing an exhaustive search on an equivalent, purely classical formulation of the original problem Hamiltonian from ~\cite{robertResourceefficientQuantumAlgorithm2021}), the actual best fits on the tetrahedral lattice (green), the actual best fits on the FCC lattice (magenta), and the best solutions from AlphaFold2 (orange).  The lattice fits and exhaustive searches were performed using a set of programs we developed.  For details on the exhaustive search algorithm, we refer the reader to App. \ref{app:ex_search}.
Following an optimization of backbone alignment, root mean square deviation, or RMSD, of the C$\alpha$ coordinates between the lattice models and a reference structure (in this case, the experimentally determined protein structure) was calculated and used to assess accuracy of each instance selected.  For each protein, the lower the value on the y-axis (RMSD), the better performing the model is.  For details on the RMSD calculations and how they were implemented, please refer to App. \ref{app:ex_search}. 
 
Fig.~\ref{bestfits} highlights a few important details.  
To begin with, in nearly every case, we see a clear gap between best solution obtained by solving the Hamiltonian of the tetrahedral lattice model (grey), and the experimental structure's best fit on the tetrahedral lattice (green).
This discrepancy we have observed highlights that further efforts must be placed in diversifying how the conformers are scored in these lattice methods, since the optimal solution does not coincide with the actual best fit on the lattice (which is closest to the experimental ground truth). 
Moreover, when comparing the actual best fits between lattices, the FCC lattice is a much more favorable choice.  
For the majority of these proteins, the best fit on the FCC lattice (magenta) has a lower RMSD than the best fit on the tetrahedral lattice (green).  
This observation supports our hypothesis: the FCC lattice is much better suited to model ordered, secondary structures of proteins, including $\alpha$-helices and $\beta$-sheets.  
In all cases, an RMSD of less than 2.0~\AA~is observed in these fits, with several instances reaching near sub-Angstrom fit (including PDBs: 2K2R~\cite{Wang2008, Wang2008_pdb}, 2MZX~\cite{Abayev2015, Abayev2015_pdb}, and 2NDE~\cite{Pastor2002, Pastor2001}).  
We also still found several cases, however, where the tetrahedral lattice fits are under 2.0~\AA. 
Lastly, we observe that for nearly half of these proteins, AlphaFold2's best models (orange) were less accurate than the best fits on either lattice based on the higher RMSD values.  
More generally, many of these AlphaFold2 models had RMSDs greater than 2.0~\AA{}, which is the common threshold used to describe an acceptable model.  
 
Figs.~\ref{bestfits}b and ~\ref{bestfits}c illustrate how these differences manifest themselves structurally.  
In both cases, it can be observed that the helix the tetrahedral lattice attempts to fit, is much more stretched out, with an average pitch between alpha carbons of roughly 9~\AA, while the FCC fit produces an average helix pitch of 6.33~\AA, in the paxillin peptide. 
Given the average pitch of 5.4~\AA found in nature, it is clear that the FCC lattice is more accurate here. 
With Trp-cage (Fig.~\ref{bestfits}b), it can also be observed that the turn and loop region is superiorly fitted by the FCC lattice as well.  
In conclusion, we empirically see in this data that the FCC lattice model is advantageous over the prior lattice models used in QPSP, especially to replicate the secondary structures observed in protein in nature.

\subsection{Comparing exhaustive search results}
\begin{figure}[t]
    \centering
    \includegraphics[scale=0.60]{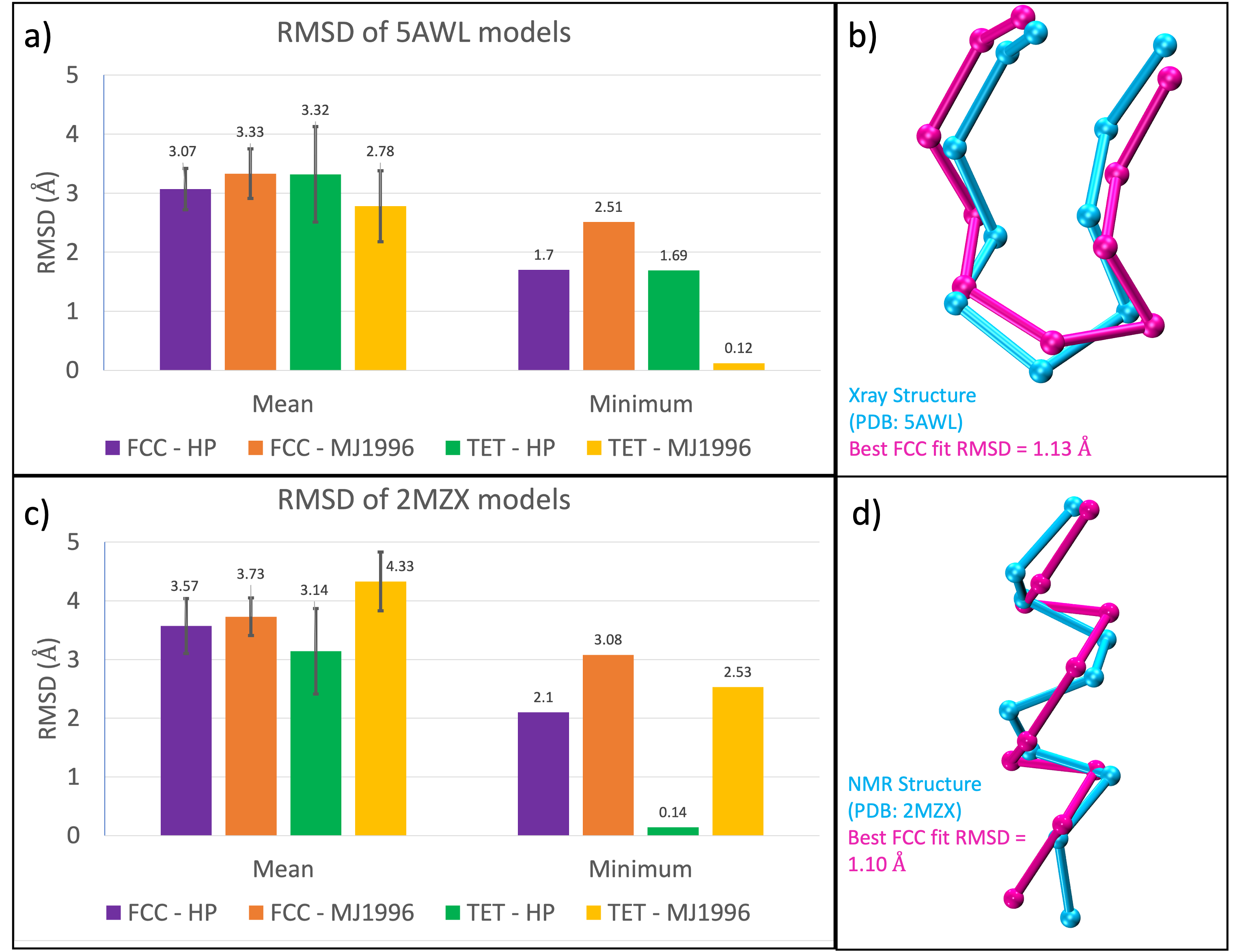}
    \caption{\small Ensemble analysis of the 100 lowest energy solutions from the exhaustive search results of the original tetrahedral lattice algorithm, and the new FCC algorithm.  Each bar plot is color coated to represent a particular lattice algorithm using a specific scoring function, either a standard hydrophobic-polar (HP) or the Miyazawa-Jernigan (MJ1996) potentials from their 1996 paper \cite{Miyazawa1996}.  For each protein structure, the mean and minimum RMSDs are with respect to the most optimal coordinates (i.e. the closest fit to the experimental structure) on each lattice.}
    \label{rmsds_chig_2mzx}
\end{figure}

To test the proposed algorithm's potential, classical exhaustive searches were performed for two peptides, each with a sequence length of 10 amino acids (See App.~\ref{app:ex_search} and Fig.~\ref{fig:rmsds_workflow} therein for further details on this process).  While this is not part of the standard workflow, it served as a meta analysis to give us a preliminary understanding of how the structures are scored on the lattice.  This is a computationally costly task that will scale poorly, however it is feasible for testing at this scale of protein for benchmarking purposes.
One of these peptides is a segment of the second extracellular loop of the C-C chemokine receptor 5 (abbreviated as CCR5 ECL2), which is known to form a helical structure in aqueous solution~\cite{Abayev2015}.  
Its primary amino acid sequence is \texttt{QYQFWKNFQT}, and it is one of the smallest helical structures in the PDB (PDB: 2MZX~\cite{Abayev2015_pdb}).  
The other peptide is a small loop structure, a mutant of ``chignolin'' (PDB: 5AWL~\cite{chig_pdb}).  
It is defined by the sequence \texttt{YYDPETGTWY}, and it is commonly used for benchmarking PSP methods ~\cite{Honda2008}.  
Together, these peptides give us an idea of how the Hamiltonians for both lattices perform in scoring loop and helical structures alike.

The exhaustive search results comprise a ranked ensemble of the 100 lowest energy structures for each case.  
Figs.~\ref{rmsds_chig_2mzx}a and c illustrate the mean and standard deviations of RMSDs in the ensemble, with respect to the optimal lattice solution, for each combination of lattice and energy function used.  
Figs.~\ref{rmsds_chig_2mzx}b and d illustrate the best fits on the FCC lattice, for each protein.  
Part of the objective of the exhaustive search is to verify if these optimal lattice solutions exist among this lowest energy ensemble.  
If the the minimum RMSD is nearly 0, the exact lattice solution was found.  This occurred once for each modeled peptide, only within the tetrahedral ensembles (TET).  
As shown in Fig.~\ref{rmsds_chig_2mzx}a, the minimum RMSD is 0.12~\AA~(yellow), for the tetrahedral lattice using the Miyazawa-Jernigan (MJ) potentials~\cite{Miyazawa1996}.  
On the other hand, in Fig.~\ref{rmsds_chig_2mzx}c we observe the minimum RMSD as 0.14~\AA, this time using the basic hydrophobic-polar (HP) model, where only first nearest neighbor pairs between hydrophobic residues are scored favorably.  
This inversion of minimum RMSDs for the TET models is also consistent with the fact that the mean RMSDs are also flipped between the modeled peptides.  
For the FCC lattice, the results are more consistent between the modeled peptides, with the HP scoring function (purple) resulting in slightly lower mean RMSDs, and a more narrow difference in the minimums, ranging from 0.8 to 1.0 \AA.  
Another thing worth noting is the difference in standard deviations (the error bars in the plot).  
The standard deviation is more narrow in the FCC versus TET models, and there is little difference when comparing between scoring functions used (0.35-0.47).  
On the other hand, there is a noticeably larger standard deviation in the TET models (0.5-0.81).  
Moreover, for the TET models, the difference in this range is more pronounced between scoring functions used per modeled peptide (0.6-0.81 for 5AWL, and 0.5-0.73 for 2MZX).  
This implies that the FCC ensembles are comprised of more similar structures in coordinate space, while the TET ensembles capture a more diverse set of structures.

This data raises several interesting questions. 
While initially surprised by the fact that the FCC ensembles don't include the optimal lattice model and only the TET ensembles do, it is actually somewhat expected.  
The added degrees of freedom in the FCC lattice versus the TET lattice (11 versus 4 possible turns per lattice node, respectively) results in a massive enlargement of the conformational space for any given protein of sequence length N.  
Statistically speaking, for the same ensemble size, the odds of finding a particular set of lattice coordinates is greater for a less `granular' lattice.  
This also, in part, helps explain the difference between the standard deviations in the RMSDs, between the ensembles.  
Following these results, to ensure the correctness of the encoding FCC lattice, we modeled and exhaustively searched the entire conformational space of a much smaller peptide (5 amino acids, sequence \texttt{GNLVS}, PDB: 4QXX~\cite{Soriaga2015, Soragni2015}).  
The exact optimal lattice model was found in this case, confirming the validity of our encoding.  
Nonetheless, one may still ask, regardless of the conformational space size between models, why wasn't the optimal lattice model \textit{scored} favorably enough to exist in the FCC's top 100 solutions?  
The question of whether all these quantum algorithms (see Table \ref{lattice_models}) are capable of optimizing in a way that yields the biological ``ground truth'' experimental structure, and the effect of the particular scoring function used, is still very much an open and rather unexplored research question in this field. 
Our team is currently investigating this in detail and our findings will be included in subsequent studies.  
In this work, we have developed a QPSP algorithm encoding the FCC itself, which, to the best of our knowledge, is the most well-suited C$\alpha$ lattice to ever be incorporated in this quantum computing application.  
It lays the ground work for future development if necessary, including enhancement of scoring methods.


\section{Solving QPSP on the face-centered cubic lattice}\label{fcc_section}

\subsection{FCC lattice encoding}

\begin{figure}[t]
    \centering
    \includegraphics[width=15cm]{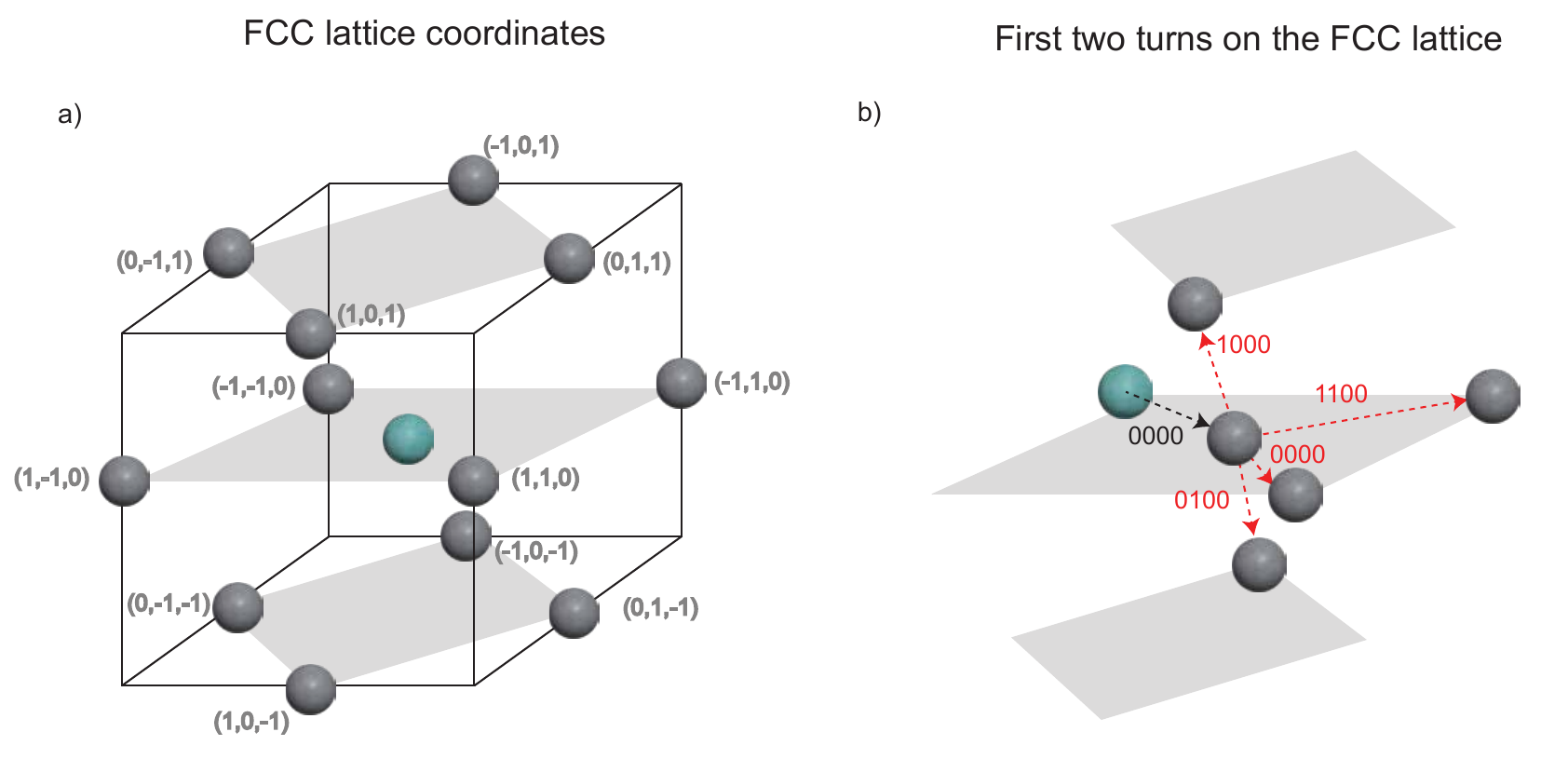}
    \caption{\small a) 12 lattice points on the FCC lattice, where the green sphere is the origin in $\mathbb{R}^3$, b) Visual depiction of the first two turns on the FCC lattice starting at the origin, and corresponding turn sequence bitstrings.}
    \label{fcc_schematic}
\end{figure}

We adopt the turn-based encoding for the QPSP problem on the FCC lattice, similar to previous works on the cubic~\cite{babejCoarsegrainedLatticeProtein2018} and tetrahedral lattices~\cite{robertResourceefficientQuantumAlgorithm2021} due to its efficiency in terms of the number of qubits required.
The FCC lattice has 12 nearest nodes surrounding each node on the lattice, resulting in a greater packing density, which allows for a more accurate representation of the protein structure than the above-mentioned lattices.
Therefore, employing the binary encoding, each turn between two consecutive amino acids can be represented by 4 qubits, resulting in $2^4 = 16$ states in total.
As shown in Fig. \ref{fcc_schematic}, the 12 distinct turn directions on the FCC lattice can be represented by 12 of these states, while the remaining 4 states are not used in the encoding.
See App.~\ref{app:encoding_details} for a detailed mapping of the turn directions to the bitstrings.

Moreover, the first turn of the protein chain can always be fixed to a specific direction because of the rotational symmetry, which reduces the number of qubits needed for the encoding.
Therefore, without loss of generality, we can assume that the first turn is in the $(1,1,0)$ direction, corresponding to the first 4 qubits being $0000$.
Even though the next turn can be any of the 12 directions, the geometry of the FCC lattice dictates that there can only be 4 unique turns, which are determined by the 4 unique angles formed by the first two turns. 
Therefore, we can fix the last two qubits in the second turn to be $00$, based on our choice of encoding.
The total qubit configuration for a length-$N$ protein/peptide is then given by
\begin{equation}
    \vb q = \underbrace{0000}_{\text{turn 0}} \underbrace{q_0q_100}_{\text{turn 1}}\underbrace{q_2q_3q_4q_5}_{\text{turn 2}}\cdots \underbrace{q_{4N-14}q_{4N-13}q_{4N-12}q_{4N-11}}_{\text{turn $N-2$}}.
\end{equation}
In total, the number of configuration qubits needed to encode the protein on the FCC lattice is $4(N-2)-2 = 4N-10$, where $N$ is the number of amino acids in the protein/peptide.

To establish the Hamiltonian for the QPSP problem, one of the fundamental building blocks for the turn-based approach is the turn indicator function, which is uniquely defined for each turn direction.
The turn indicator function is a binary function that takes the qubit configuration as input and outputs 1 if the configuration corresponds to the given turn direction, and 0 otherwise.
For example, the indicator function for the $(1,1,0)$ turn, represented by 0000, is given by
\begin{equation}
    d^j_{(+x,+y)} = (1-q_{\phi+2})(1-q_{\phi+3})(1-q_{\phi+4})(1-q_{\phi+5}),
\end{equation}
where $\phi = 4(j-2)$, and $j=2, \cdots, N-2$ is the index of the turn.
The indicator functions for the other 11 turns can be similarly defined.
Additionally, we will also make use of indicator functions for the 4 unused turns that do not correspond to any physical turn: $d^j_{0010}$, $d^j_{0001}$, $d^j_{1101}$, and $d^j_{1110}$.
We summarize the explicit expressions of the turn indicator functions in App.~\ref{app:encoding_details}.

With the turn indicator functions, we can now define the Hamiltonian for the protein folding problem on the FCC lattice.
In this case, the Hamiltonian consists of four terms~\cite{babejCoarsegrainedLatticeProtein2018}:
\begin{equation}
    H(\vb q) = H_{\text{back}}(\vb q) + H_{\text{redun}}(\vb q) + H_{\text{olap}}(\vb q) + H_{\text{int}}(\vb q).
\end{equation}
The first term $H_{\text{back}}$ imposes the non-backtracking constraint, which penalizes the configurations that backtrack to the previous turn, i.e., cases where any two consecutive turns are in opposite directions.
The second term $H_{\text{redun}}$ penalizes the occurrence of the four redundant bitstrings that do not correspond to any physical turn.
The third term $H_{\text{olap}}$ penalizes the overlap between any two amino acids in the protein, which ensures that each amino acid occupies a unique position on the lattice.
Finally, the last term $H_{\text{int}}$ is the interaction term that accounts for the nearest-neighbor interactions between any two amino acids.
Below we will illustrate in detail how each term is constructed.

\subsection{Hamiltonian construction}
\subsubsection*{Non-backtracking constraint}
The construction of $H_{\text{back}}$ is straightforward and has been discussed extensively in the literature~\cite{babbushConstructionEnergyFunctions2014, babejCoarsegrainedLatticeProtein2018, robertResourceefficientQuantumAlgorithm2021}.
Basically, it relies on the pseudo-Boolean expressions of the turn indicator functions, which can be used to construct the penalty terms for the non-backtracking constraint.
For example, we want to penalize two consecutive turns that are in the $(+x, +y)$ and $(-x, -y)$ directions, respectively, and vice versa.
Such situation can be captured by the following expression,
\begin{equation}
    \begin{split}
     & \qty[d^j_{(+x,+y)}\land d^{j+1}_{(-x,-y)}] \lor \qty[d^j_{(-x,-y)}\land d^{j+1}_{(+x,+y)}] \\  
     = \hphantom{\ } & (1-q_{4j-6})(1-q_{4j-5})(1-q_{4j-4})(1-q_{4j-3}) \cdot (1-q_{4j-2})(1-q_{4j-1})q_{4j}q_{4j+1} \\
     & + (1-q_{4j-6})(1-q_{4j-5})q_{4j-4}q_{4j-3} \cdot (1-q_{4j-2})(1-q_{4j-1})(1-q_{4j})(1-q_{4j+1}),
    \end{split}
\end{equation}
where $\land$ and $\lor$ denote the logical AND and OR operations, respectively.
It is worth noting that the above expression will result in Hamiltonian terms that have a maximum locality of 8.
The non-backtracking constraint can be enforced by summing up all such terms for all possible pairs of consecutive turns along the protein chain.
The resulting Hamiltonian can be written as
\begin{equation}
    \begin{split}
        H_{\text{back}}(\vb q) = &\ \lambda_\text{back} \Big[(1-q_0)(1-q_1) \land d^2_{(-x,-y)} + (1-q_0)q_1 \land d^2_{(+x,+z)} \\
        & + q_0(1-q_1) \land d^2_{(-x,-z)} + q_0q_1 \land d^2_{(+x,-y)} + \sum_{j=2}^{N-3} \sum_{a} d^j_{a}\land d^{j+1}_{-a} \Big],
    \end{split}
\end{equation}
where $\lambda_\text{back}$ is the penalty parameter that controls the strength of the non-backtracking constraint, and $a \in \{0, \dots, 11\}$ denotes the 12 distinct turn directions.

\subsubsection*{Redundant bitstrings}
To avoid the occurrence of the four unphysical bitstrings, we can similarly construct the penalty terms based on the turn indicator functions.
Essentially, whenever any of the four redundant bitstrings appear in the configuration, we want to penalize it.
Therefore, the Hamiltonian can be written as
\begin{equation}
    H_\text{redun}(\vb q) = \lambda_\text{redun} \sum_{j=2}^{N-2} \qty[d^j_{0010} + d^j_{0001} + d^j_{1101} + d^j_{1110}],
\end{equation}
with a different penalty parameter $\lambda_\text{redun}$.

\subsubsection*{Nearest-neighbor interactions}
Furthermore, interactions between any non-covalently bonded neighboring amino acids on the lattice need to be accounted for.
This is truly what is driving the protein folding process, as the amino acids interact with each other through non-covalent forces such as van der Waals interactions, hydrogen bonds, and hydrophobic interactions.
For each possible interaction pair, we need to add one ancilla qubit flag $q^a_{mn}$ that is equal to 1 if the pair are nearest neighbors, and 0 otherwise.
To determine the nearest neighbors on a lattice, we need to be able to keep track of the relative distances between all pairs of amino acids.
To do this, we first attain the coordinates of each amino acid in the protein sequence.
They are calculated using the position functions $(x_m, y_m, z_m)$, with $m > 0$ denoting the amino acid index (see App.~\ref{app:encoding_details}).
Subsequently, we can define the squared distance function as
\begin{equation}
    D_{mn} = (x_m - x_n)^2 + (y_m - y_n)^2 + (z_m - z_n)^2.
\end{equation}
On the FCC lattice, a pair of beads $m$ and $n$ are deemed nearest neighbor if $D_{mn} = 2$.
While the construction of the interaction Hamiltonian is straightforward, it is worth pointing out that unlike the cubic or tetrahedral lattice, amino acids separated by an even number of turns on the FCC lattice can be nearest neighbors due to the non-bipartite nature of it.
To this end, the interaction Hamiltonian can be written as
\begin{equation}
    H_\text{int}(\vb q) = \sum_{m=0}^{N-3}\sum_{n=m+2}^{N-1} q^a_{mn}\epsilon_{mn}(3 - D_{mn}),
\end{equation}
where $\epsilon_{mn}$ is the interaction strength between the amino acids at positions $m$ and $n$, which can be determined from a contact energy matrix Miyazawa-Jernigan potentials.  These contact energies are widely used in these applications and represent a statistical analysis of pairing frequencies between amino acids in a database of proteins, which is intended to approximate the free energy associated with these interactions.  Thus, the interaction Hamiltonian serves as a scoring function, allowing us to differentiate between optimal and suboptimal nearest neighbor pairs.

\subsubsection*{Non-overlapping constraint}

\begin{figure}[t]
    \centering
    \includegraphics[width=0.55\textwidth]{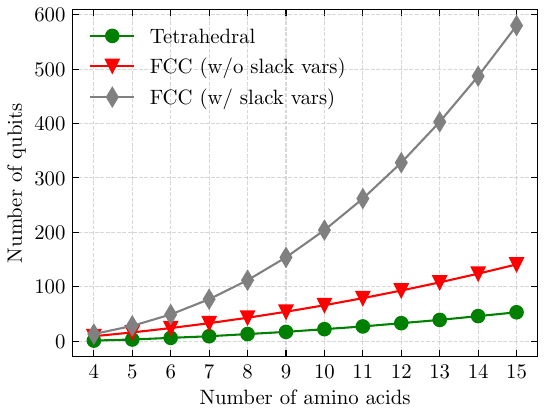}
    \caption{\small Qubit counts vs.~protein length for the FCC with slack variables (grey), without slack variables (red), and tetrahedral lattice (green) implementations of the QPSP problem.}
\label{fig:qubit_counts}
\end{figure}

Finally, in order to penalize the potential overlaps between any two amino acids on the lattice that are not necessarily adjacent, we can make use of the same squared distance function introduced above.
Mathematically, the non-overlapping constraint requires that $D_{mn} > 0$ for all $m \neq n$.
Furthermore, the geometry of the FCC lattice gives an upper bound on the squared distance.
Therefore, we arrive at the following inequality constraints:
\begin{equation}
    2 \leq D_{mn} \leq 2(m-n)^2,
\end{equation}
where the lower bound corresponds to the minimum distance allowed between any two amino acids on the lattice, that is, the nearest neighbor distance.
In the existing literature, such inequality constraints are enforced by introducing the so-called slack variables, which are non-negative and penalize the violation of the constraints ~\cite{babbushConstructionEnergyFunctions2014, babejCoarsegrainedLatticeProtein2018}. 
This allows us to convert the inequality constraint into an equality constraint, which can then be added to the Hamiltonian as a penalty term as before.
However, the encoding of the slack variables requires additional ancilla qubits for each pair $(m, n)$, which in total scale as $\mathcal{O}(N^2\log N)$ with the protein length $N$.
In Fig.~\ref{fig:qubit_counts}, we show the total qubit count as a function of the protein length $N$ for the FCC lattice with and without using slack variables.
We also show a direct comparison with the tetrahedral lattice case - while the FCC lattice certainly requires more qubits than the tetrahedral one, asymptotically both require $\mathcal{O}(N^2)$ qubits.
As can be seen in Fig.~\ref{fig:qubit_counts}, the introduction of slack variables significantly increases qubit counts, which severely limits the length of the protein sequences that can be solved on pre-fault tolerant quantum computers.

In the following subsections, we present two approaches that avoid using slack variables to construct the non-overlapping Hamiltonian $H_\text{olap}$.

\subsection{Alternative approaches to the non-overlapping constraint}

\subsubsection{Fitting with higher order polynomials} \label{sec:polyfit}

\begin{figure}[t]
    \centering
    \includegraphics[width=0.9\textwidth]{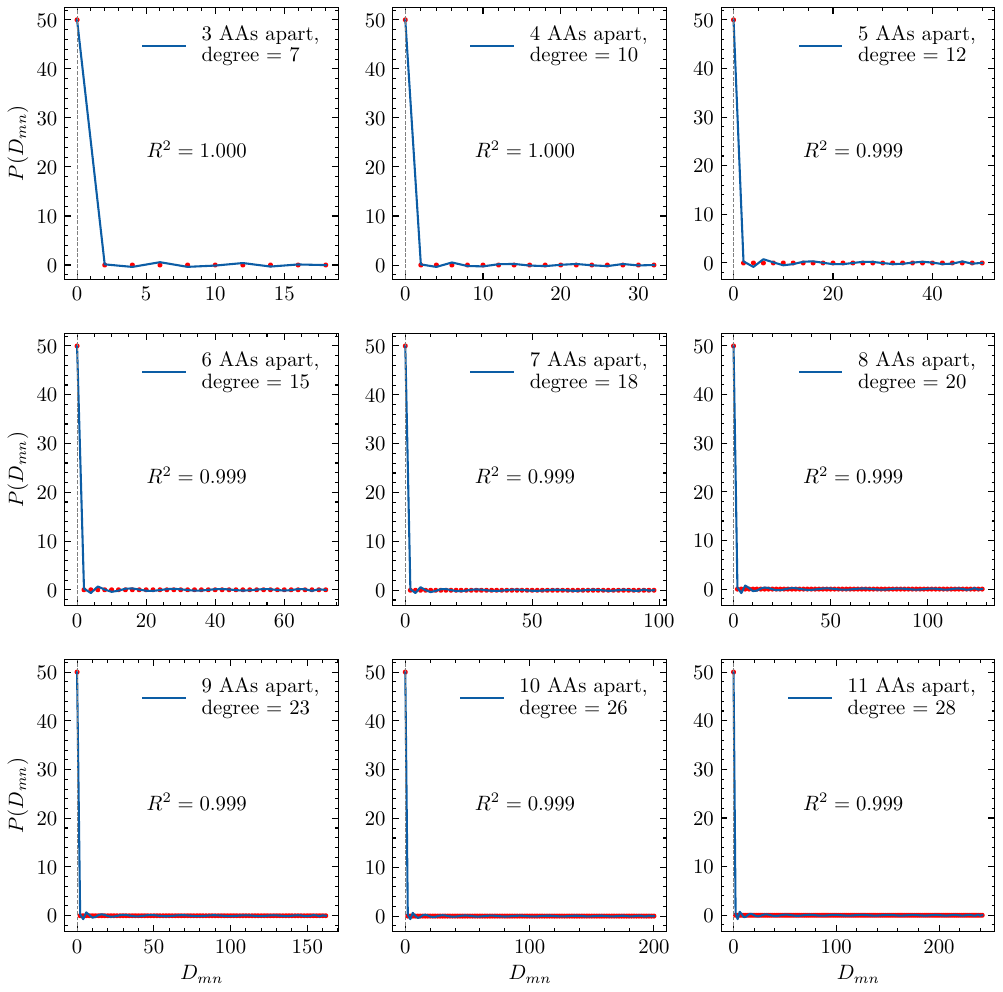}
    \caption{\small Polynomial fits to the desired values of the penalty term $F[D_{mn}]$ (corresponding to a penalty parameter of 50 when bead overlap happens) for the non-overlapping constraint. Cases where $\abs{m - n} = 3, \dots, 11$ are presented. The polynomial fit is obtained by minimizing the least squares error for a given tolerance ($R^2 = 0.999$ in this case). The degree of the polynomial is determined through Algorithm~\ref{alg:polyfit_math}.}
\label{fig:polyfit}
\end{figure}

One way to potentially avoid the use of slack variables in encoding inequality constraints is the unbalanced penalization method, which was introduced in Ref.~\cite{montanez-barreraUnbalancedPenalizationNew2023} for quadratic optimization problems.
The unbalanced penalization approach introduces a quadratic approximation to an exponential filter functional $F[h(x)] = e^{-h(x)}$ for the inequality constraint $h(x) > 0$.
Ideally, with the exponential filter functional, the penalty is close to 0 for positive values of $h(x)$, while for negative values, it exponentially increases, thus enforcing the constraint.
Approximating the exponential filter with a quadratic function makes the Hamiltonian construction easier, as it can be expressed in terms of Pauli strings.
However, in practice, especially for the QPSP problem, where the energy hierarchy of different configurations is sensitive to the approximation error, a quadratic penalty functional may not be sufficient to capture the desired behavior of the penalty term.
In particular, the quadratic penalty functional tends to introduce large biases even at allowed values of $D_{mn}$, which can lead to suboptimal solutions.

To overcome this issue, we consider incorporating higher order terms in the penalty functional $F[D_{mn}]$.
An ideal penalty functional should have a large, positive value at $D_{mn} = 0$ and zero at all other possible values of $D_{mn}$, $\{2, 4, \dots, 2(m-n)^2\}$.
To find such a functional, we may consider fitting a polynomial to the desired values of $F[D_{mn}]$ through the method of least squares for a given tolerance.
The tolerance can be chosen based on the desired accuracy of the penalty term, e.g., $R^2$ value of the fit.
This would in turn determine the minimum degree of the polynomial for the fit.
We make use of the Chebyshev polynomial basis to construct the polynomial fit to the desired values of $F[D_{mn}]$.
The algorithm for the higher order polynomial fit is summarized in Algorithm~\ref{alg:polyfit_math}.

As shown in Fig.~\ref{fig:polyfit}, for a given tolerance, the minimum degree of the polynomial is determined by the number of possible values of $D_{mn}$, which is in turn determined by the maximum distance between any two amino acids on the lattice.
The further the amino acids are from each other, the higher the degree of the polynomial required to construct the penalty term.
One can see that the degree of the polynomial grows roughly linearly with the number of amino acids in the protein chain.
This poses a significant challenge for the problem formulation in this approach, as the number of Hamiltonian terms grows exponentially with the degree of the polynomial.
Overall, as the peptide length increases, the number of Hamiltonian terms grows exponentially, which can lead to an exponential growth in the classical computation time required to estimate the expectation values of the Hamiltonian.
As a result, the number of amino acids that can be simulated with such an approach can be severely limited.
Alternatively, one may limit the maximum degree of the polynomial to a fixed value depending on the available computational resources, which would lead to a suboptimal solution.
However, for practical purposes, such an approximate solution may still be sufficient for, e.g., the initial screening of the protein conformational space.
Finding a suitable balance between the polynomial degree and the accuracy of the penalty term is an important aspect of this approach and we leave it for future work.

\begin{algorithm}[t]
\caption{Polynomial fit to the target values of the penalty term $F[D_{mn}]$}
\label{alg:polyfit_math}
\begin{algorithmic}[1]
\State \textbf{Given:} Data points $\{(D_i, F_i)\}_{i=1}^M$, tolerance $R^2_{\mathrm{tol}}$
\State \textbf{Goal:} Find coefficients $\{a_k\}_{k=0}^d$ such that
\[
P_d(D) = \sum_{k=0}^d a_k T_k(D), \quad R^2(P_d) \geq R^2_{\mathrm{tol}}
\]
\State Initialize polynomial degree $d \gets d_0$
\While{$R^2(P_d) < R^2_{\mathrm{tol}}$}
    \State $d \gets d + 1$
    \State Fit $P_d(D)$ to $\{(D_i, F_i)\}$ using least squares in the Chebyshev basis
    \State Compute $R^2(P_d)$
\EndWhile
\State \textbf{Return:} Polynomial coefficients $\{a_k\}_{k=0}^{d}$
\end{algorithmic}
\end{algorithm}

\subsubsection{Lagrangian duality} \label{sec:lagrangian_duality}

One promising approach to avoid the above-mentioned exponential complexity of the PolyFit while still enforcing the non-overlapping constraint without slack variables is the \textit{Lagrangian duality}. 
In general, Lagrangian duality transforms a constrained optimization problem into an unconstrained one by incorporating the constraints into the objective function using weighted penalty terms called dual variables. 
However, instead of adding the penalty terms directly to the Hamiltonian, we can use the Lagrangian duality to reformulate the problem in a way that allows us to optimize the dual variables separately from the primal variables.
Below we briefly outline the main ideas of the Lagrangian duality and how it can be applied to the QPSP problem on the FCC lattice.
Starting with a typical binary optimization problem with inequality constraints:
\begin{alignat}{3}
P^{*} & := \min_{\vb b\in\{0,1\}^n} && f_{0}(\vb b) \\
& \hphantom{==} \text{subj. to}\ && f_{m}(\vb b)\leq 0, \quad m = 1:M, 
\end{alignat}
where $f_0$ is the objective function and $f_m$ are the inequality constraints, e.g., the non-overlapping constraints in our problem.
$\vb b \in \{0, 1\}^n$ denotes length-$n$ binary strings (bitstrings). 
This is known as the \textit{primal} problem.
Introducing the \textit{Lagrange multipliers} $\lambda_m$, we can define the Lagrangian function as
$$
\mathcal L(\vb b; \boldsymbol{\lambda}) := \sum_{m=0}^M \lambda_m f_m(\vb b),
$$
where $\lambda_0 = 1$. 
One can show that the original (primal) problem is equivalent to 
$$
\min_{\vb b} \max_{\boldsymbol{\lambda}\succeq 0} \mathcal L(\vb b; \boldsymbol{\lambda}).
$$
Reversing the order of min and max, we can define the associated dual function as
$$
D(\boldsymbol{\lambda}) := \min_{\vb b} \mathcal L(\vb b; \boldsymbol{\lambda}),
$$
and the corresponding \textit{Lagrangian dual} problem then aims at maximizing $D$:
$$
D^* = \max_{\boldsymbol{\lambda}\succeq 0} D(\boldsymbol{\lambda}).
$$
Note that the optimal solution to the dual problem is a lower bound for the optimal solution of the original (primal) problem, i.e., $D^* < P^*$; this is known as the \textit{weak duality} principle.

Recently, \citet{leVariationalQuantumEigensolver2024} adapted the Lagrangian duality method to the variational quantum eigensolver (VQE) algorithm for quantum optimization, which they named the variational quantum eigensolver with constraints (VQEC).
In the context of VQE, the objective function and the constraints are given as the expectation values of the corresponding operators:
\begin{equation*}
    f_m(\boldsymbol{\theta}) = \ev{H_m}{\psi(\boldsymbol{\theta})}, \quad m = 0:M,
\end{equation*}
where $H_0$ is the Hamiltonian of the problem, and $H_m$ ($m \geq 1$) are the constraint operators corresponding to the inequality constraints.
For our problem, $H_m$ represent inequalities for the non-overlapping constraints, i.e., $2 - D_{mn}$ for all possible pairs of amino acids $m$ and $n$.
Therefore, the number of constraints $M$ scales as $\mathcal{O}(N^2)$ with the protein length $N$.
$\ket{\psi(\boldsymbol{\theta})}$ is the quantum state parameterized by $\boldsymbol{\theta}$, i.e., the output of the variational quantum circuit.
Therefore, the primal problem formulated in the VQE framework becomes
\begin{alignat*}{3}
    P^{*}_\theta & := \min_{\boldsymbol{\theta}} && f_{0}(\boldsymbol{\theta}) \\
    & \hphantom{==} \text{subj. to}\ && f_{m}(\boldsymbol{\theta})\leq 0, \quad m = 1:M, 
\end{alignat*}
while the dual problem becomes 
\begin{equation*}
    D_\theta^* = \max_{\boldsymbol{\lambda}\succeq 0} \min_{\boldsymbol{\theta}} \mathcal L_\theta(\boldsymbol{\theta}; \boldsymbol{\lambda}).
\end{equation*}
To adequately solve the dual problem using VQEC, the authors employed the primal-dual perturbation (PDP) method~\cite{leVariationalQuantumEigensolver2024, kallioPerturbationMethodsSaddle1994, kallioLargeScaleConvexOptimization1999}. 
At each step of VQE iteration, one first perturbs the primal and dual variables as
\begin{align*}
\tilde{\boldsymbol{\theta}}^t &= \qty[\boldsymbol{\theta}^t - \nu_\theta \sum_{m=0}^M \lambda_m^t \nabla_{\boldsymbol{\theta}} F_m(\boldsymbol{\theta}^t)]_\theta, \\
\tilde{\lambda}_m^t &= \left[\lambda_m^t + \nu_\lambda F_m(\boldsymbol{\theta}^t)\right]_+, \qquad m = 1:M,
\end{align*}
where $[x]_\theta$ denotes the projection of $x$ onto the feasible range of $\boldsymbol{\theta}$, e.g., $\boldsymbol{\theta} \in [0, 2\pi]^P$ for single-qubit rotations,  $[x]_+ := \max\{x, 0\}$.
$\nu_\theta$ and $\nu_\lambda$ are perturbation step sizes. 
Once the perturbed variables are generated, one updates the primal variables via gradient descent and the dual variables via gradient ascent:
\begin{align*}
\boldsymbol{\theta}^{t+1} &= \qty[\boldsymbol{\theta}^t - \mu_\theta^t \sum_{m=0}^M \tilde{\lambda}_m^t \nabla_{\boldsymbol{\theta}} F_m(\boldsymbol{\theta}^t)]_\theta, \\
\lambda_m^{t+1} &= \left[\lambda_m^t + \mu_\lambda^t F_m(\tilde{\boldsymbol{\theta}}^t)\right]_+, \qquad m = 1:M,
\end{align*}
where $\mu_\theta^t$ and $\mu_\lambda^t$ are the step-dependent update step sizes. 
It has been shown in Ref.~\cite{kallioPerturbationMethodsSaddle1994} that with carefully chosen step sizes, the primal-dual iterates can converge to a saddle point of the $\mathcal L$, without any strict convexity assumptions on $\mathcal L$.

In summary, the approach allows us to solve the constrained optimization problem without explicitly constructing the Hamiltonian terms for the constraints, which is a significant advantage over the polynomial penalty method.
However, it comes with the cost of needing to run more quantum circuits per iteration due to the gradient evaluations.
Assuming the use of the parameter-shift rule~\cite{schuld2019evaluating}, the total number of quantum circuits required for the PDP method is $2(P+1)$ per iteration, where $P$ is the number of parameters in the quantum circuit~\cite{leVariationalQuantumEigensolver2024}.
Moreover, the PDP method introduces additional complexity in the optimization process, as one needs to carefully tune the various step sizes to ensure convergence.
More theoretical and empirical studies on the duality and optimality gaps of the Lagrangian duality method are needed to better understand the behavior of the VQEC algorithm and to develop more efficient optimization strategies.
As will be shown in the Sec.~\ref{sec:sim-results}, in our experiments, we perform a grid search over the various hyperparameters of the VQEC algorithm, including the perturbation and update step sizes.
For each set of hyperparameters, we run the optimization multiple times without random initial primal variables to account for stochasticity in the optimization process.
We then select the best-performing hyperparameters based on the convergence behavior of the optimization process and the quality of the final solution.
The parameters from the best-performing optimization run are then used to generate the final results on the quantum computer.

\subsection{Quantum optimization of the FCC Hamiltonian} \label{sec:optimization}
In this subsection, we present more details on the quantum optimization of the FCC Hamiltonian for the QPSP problem.
To begin with, we need to choose the quantum circuit ansatz for the variational optimization.
In general, two types of quantum circuits are commonly used for VQAs: the problem-specific ansatze and the hardware-efficient ones.
A problem-specific ansatz is tailored to the specific problem at hand.
A well-known example of such an ansatz in quantum optimization is the quantum approximate optimization algorithm (QAOA) ansatz \cite{farhi2014quantumapproximateoptimizationalgorithm}.
The QAOA ansatz requires one to embed the problem Hamiltonian into a quantum circuit, where the terms of the Hamiltonian are represented by quantum gates.
The complexity of the gates in the QAOA ansatz is largely determined by the number of non-trivial (non-identity) Pauli operators in the Hamiltonian terms.
For example, a term like $ZZ$ is mapped to the $R_{ZZ}$ gate, which is a two-qubit gate that can be decomposed into a single-qubit $R_Z$ gate and two CNOT gates.
A higher-order $ZZZ$ term, on the other hand, is mapped to a three-qubit gate that decomposes into a single-qubit $R_Z$ and four CNOTs.
As the locality of the Hamiltonian terms increases, the number of gates and circuit depth also increase, which can become challenging for current quantum hardware.
On top of that, non-local interactions in the Hamiltonian can further exacerbate the problem, as they require routing of the qubits if the hardware does not have a native connectivity that matches the problem Hamiltonian.
Given that the FCC Hamiltonian has a large number of non-local and high-order interactions, especially when it is generated with the PolyFit approach (see Sec.~\ref{sec:polyfit}), it poses significant challenges for the QAOA ansatz, or any problem-specific ansatz in general, which can lead to a large number of gates and high circuit depth.
On the other hand, a hardware-efficient ansatz (HEA) is designed to be compatible with the qubit topology of the quantum hardware, which allows for a more efficient execution on the quantum computer \cite{kandala2017hardware}.
It has shown some promise in solving the QPSP problem on the tetrahedral lattice in noisy simulations, albeit assuming all-to-all connected qubits~\cite{robertResourceefficientQuantumAlgorithm2021}.
It also provides a more flexible and adaptable framework for different problems, as it does not require the explicit embedding of the problem Hamiltonian into the circuit.
As such, with our current formulation, the HEA is the only viable option for executing the QPSP workflow on current quantum devices.
Specifically, we choose the \texttt{RealAmplitudes} ansatz~\cite{RealAmplitudesLatestVersion} with 2 layers for our experiments, which is a hardware-efficient ansatz that contains a sequence of single-qubit $R_Y$ rotations and CNOT gates in its repeated structure.


\section{Results}

In this section, we present and discuss the results we obtained by implementing the FCC lattice model for the PSP problem on quantum computers, the first time in the field to the best of our knowledge, along with the simulation results for training the ansatz to determine the optimal parameters.
We focused on modeling a short, 6-residue peptide (\texttt{KLVFFA}, PDB:2Y29 \cite{Colletier2010}) which is part of a specific polymorph of the amyloid-beta (A$\beta$) peptide sequence \cite{Colletier2011}. 
These A$\beta$ peptides in turn aggregate into $\beta$-sheets, forming the plaques in the brain that are commonly associated with the pathogenesis of Alzheimer's disease.  
Using the encoding and Hamiltonian construction introduced in Sec.~\ref{fcc_section}, this 6-residue problem maps to 24 qubits.
Thus, the ensembles we predicted with the new FCC algorithm on quantum hardware, represent a use case that is both quantum computationally efficient and biomedically significant.

\subsection{Simulation results}\label{sec:sim-results}

For both methods, PolyFit and VQEC, we first run the training routines described below and identify the best parameters for the ansatz to recover the optimal solution.   

For the PolyFit approach, we adopt the Conditional Value-at-Risk (CVaR) optimization strategy~\cite{barkoutsos2020improving} in the VQE algorithm, where the objective function is defined as the average of the low-energy tail of the energy distribution, delimited by a threshold value $\alpha$.
The CVaR-VQE allows us to focus on the low-energy solutions that are relevant for the protein folding problem, while also providing a speedup in the optimization process.
In our case, we set the threshold value $\alpha = 0.1$.
The optimizer used in our experiments is the COBYLA (Constrained Optimization BY Linear Approximations) algorithm~\cite{powellDirectSearchOptimization1994}, which is a derivative-free optimizer that is less susceptible to local minima in the optimization landscape.
To reduce the stochasticity in the parameter initialization and the optimization process, we run the optimization multiple times with different random initial parameters and select the best-performing optimization run based on the convergence behavior and the quality of the final solution. To find the optimal parameters for convergence to the ground state solution, we have divided the parameter space into four subintervals, and sampled the rotational angles from these subintervals of $[0,2\pi]$. Observing that the most successful initial parameters are obtained from the interval $[\frac{3\pi}{2} , 2\pi]$, we have repeated sampling from this subinterval until we have identified the best convergence properties for the optimizer in terms of the number of classical optimizer iterations needed and the success probability of recovering the best solutions.

In the case of the VQEC approach, we use the PDP method described in Sec.~\ref{sec:lagrangian_duality} to optimize the dual variables and the primal variables simultaneously.
In this case, the objective function is defined as simply the expectation value of the Hamiltonian and the gradients are computed via the parameter-shift rule~\cite{schuld2019evaluating}.
We initialize all the dual variables to zero and the primal variables to random values within the feasible range, i.e., $[0, 2\pi]$ for single-qubit rotations.
We additionally perform a hyperparameter search over the various values of the perturbation and update step sizes to identify the best-performing run of the VQEC algorithm.
For simplicity, we use the same perturbation step size for both the primal and dual variables, i.e., $\nu_\theta = \nu_\lambda$.
Same update step sizes are also used for both primal and dual variables, i.e., $\mu_\theta = \mu_\lambda$.
The perturbation step sizes tested in our experiments are $\nu \in  \{0.01,~0.05,~0.1,~0.2,~0.5\}$, and the update step sizes are $\mu \in \{0.1,~0.5,~1,~2,~5\}$.
For each set of hyperparameters, we run the optimization 20 times with different initial primal variables.

\begin{figure}[htbp]
    \centering
    \includegraphics[width=\textwidth]{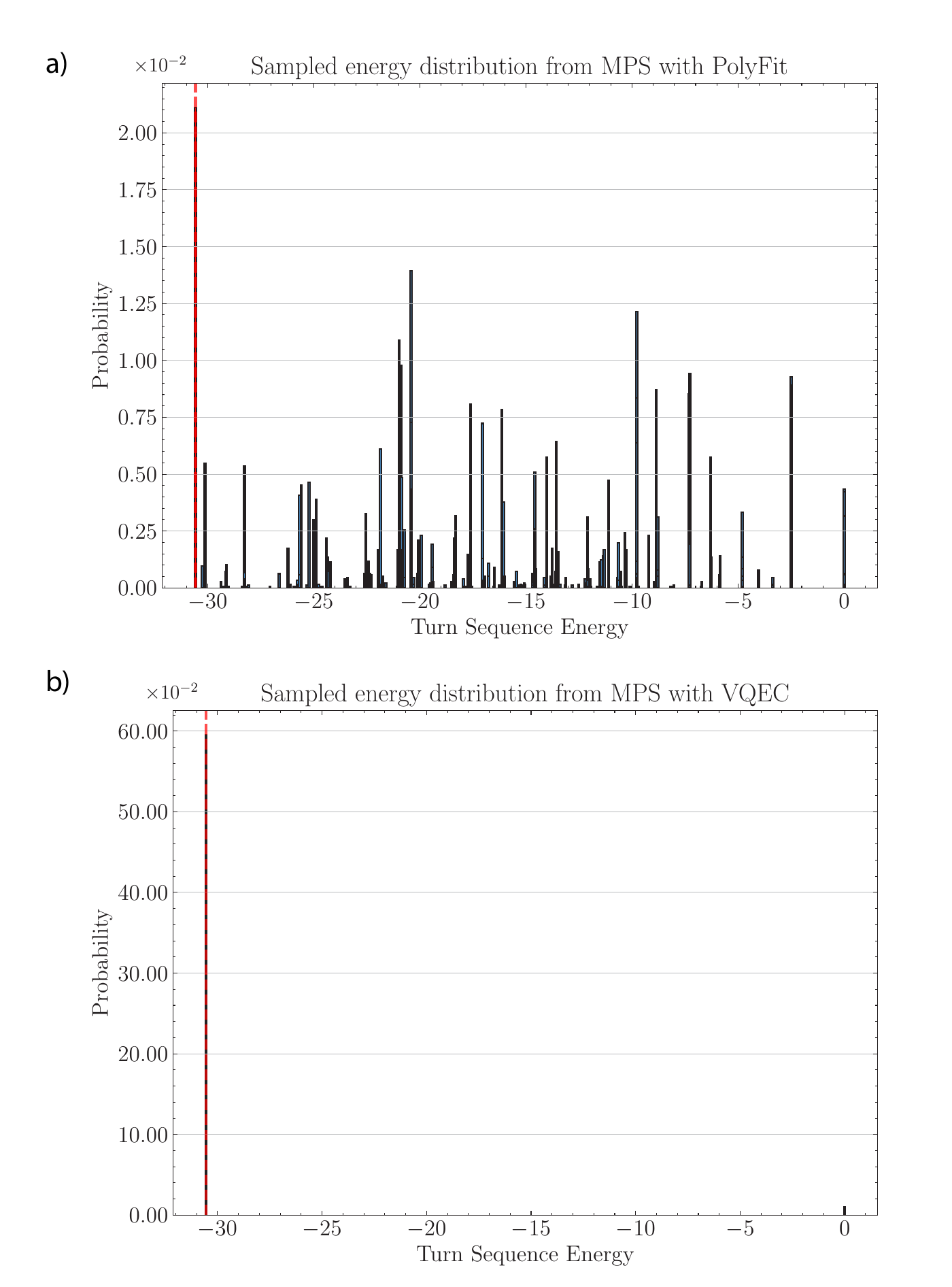}
    \caption{\small 
    Turn sequence probabilities for the protein sequence \texttt{KLVFFA} when using the a) PolyFit method and b) VQEC method. The optimized circuit is sampled with 100,000 shots on a noiseless simulator and the sampled bitstrings are subsequently converted to the corresponding turn sequences. The red dashed line indicate the ground state energy found by the classical exhaustive search.}
    \label{mps-plot}
\end{figure}

The optimized circuits are then sampled on a noiseless simulator, where the bitstrings and their probabilities are recorded (See Fig.~\ref{mps-plot}) to benchmark the hardware experiments. 
We can see that with either method, we are able to identify the parameters that yield the ground state solutions. 
It is worth noting that there is a stark difference between the sampling distributions obtained from the PolyFit and VQEC approaches.  
Although the PolyFit parameters recover a ground state solution with the highest probability among the distribution, this probability remains relatively low.
Furthermore, a wide spectrum of suboptimal bitstrings are sampled with a comparable degree of probability (similar to what is observed in the hardware experiments).  
On the other hand, the sampling with the optimal VQEC parameters recovers the ground state with nearly 60\% probability (almost 30 times that of the PolyFit approach), with almost no sampling of the suboptimal bitstrings. 
This suggests that the VQEC method is able to effectively optimize the quantum circuit parameters to recover the ground state solution with high probability, while the PolyFit method is more prone to sampling suboptimal solutions.
While we clearly observe the advantage of the VQEC method and its ability to consistently sample the best solution, it also requires a meticulous fine-tuning of the hyper-parameters as mentioned earlier. 
Classical post-processing is then performed to extract the turn sequences and subsequently the ensembles of the predicted protein conformations.
We inspect the sampled conformations to check for overlaps and other violations of the constraints, and compare against the ground truth conformations obtained from the classical exhaustive search.

The best-performing run is selected by evaluating which sampled solutions recover the top conformations (with lowest energies), and by comparing the probabilities assigned to these optimal solutions. These parameters are later used in actual quantum computer executions.

 \subsection{Hardware results}\label{sec:hardware-results}
 
\begin{figure}[htbp]
    \centering
    \includegraphics[width=\textwidth]{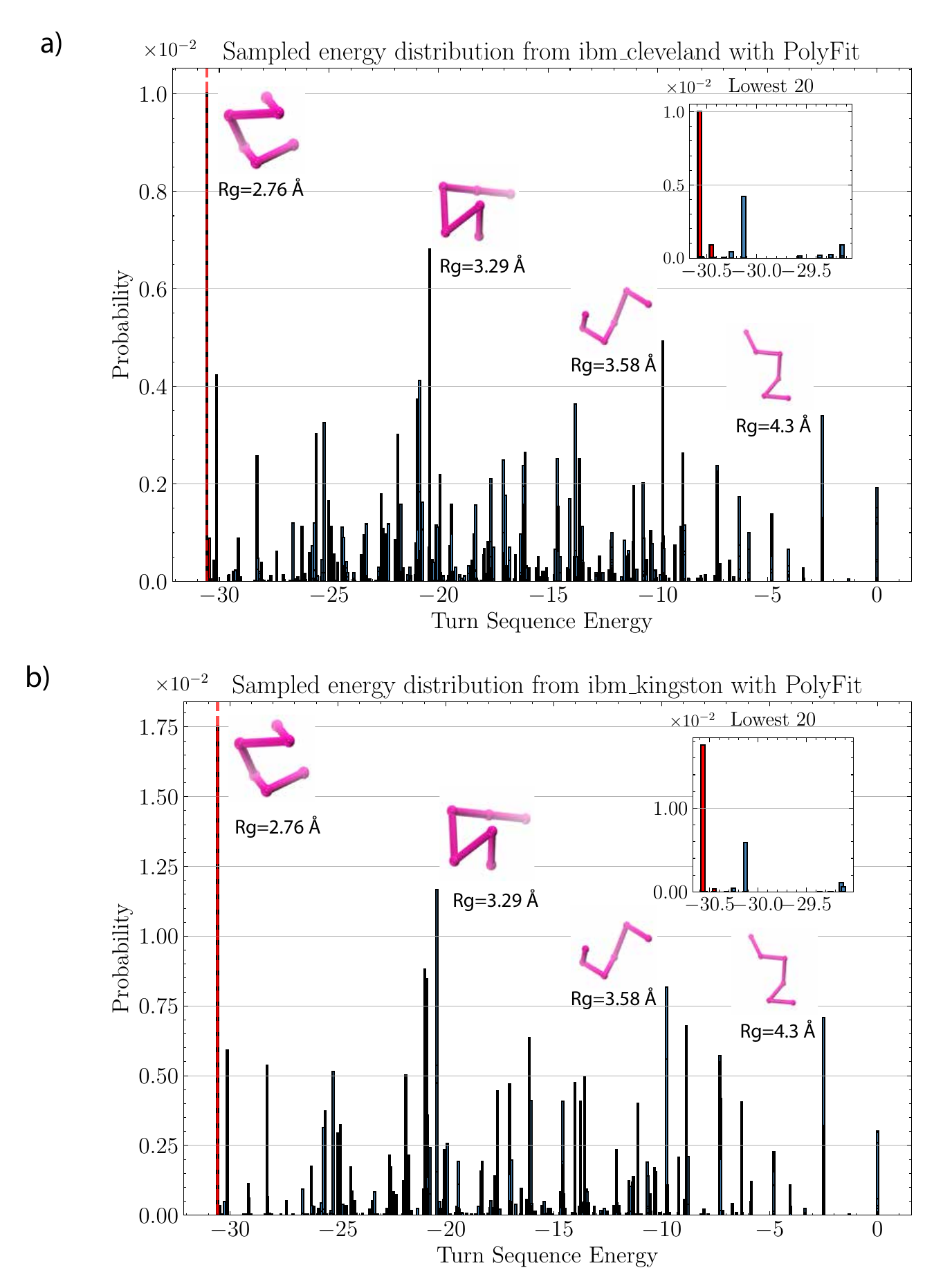}
    \caption{\small Turn sequence probabilities for \texttt{KLVFFA} ensembles when using the PolyFit approach, on a) \texttt{ibm\_cleveland} and b) \texttt{ibm\_kingston}. The red dashed line indicate the ground state energy found by the classical exhaustive search. In both machines, we observe at least one of the ground state turn sequence bitstrings having the highest probability within the distribution, while presenting other clusters of high probability, high energy bitstrings throughout the distribution. The same four predominant bitstrings are observed in both plots, with gradually increasing probabilities. Radius of gyration is computed for the associated structures, with noticeable reduction in these values as probabilities increase and energy decreases. On average, for the same bit strings, up to a nearly 2-fold increase in probability is seen for the results from \texttt{ibm\_kingston} versus \texttt{ibm\_cleveland}. }
    \label{cheb-plot}
\end{figure}

\begin{figure}[htbp]
    \centering
    \includegraphics[width=\textwidth]{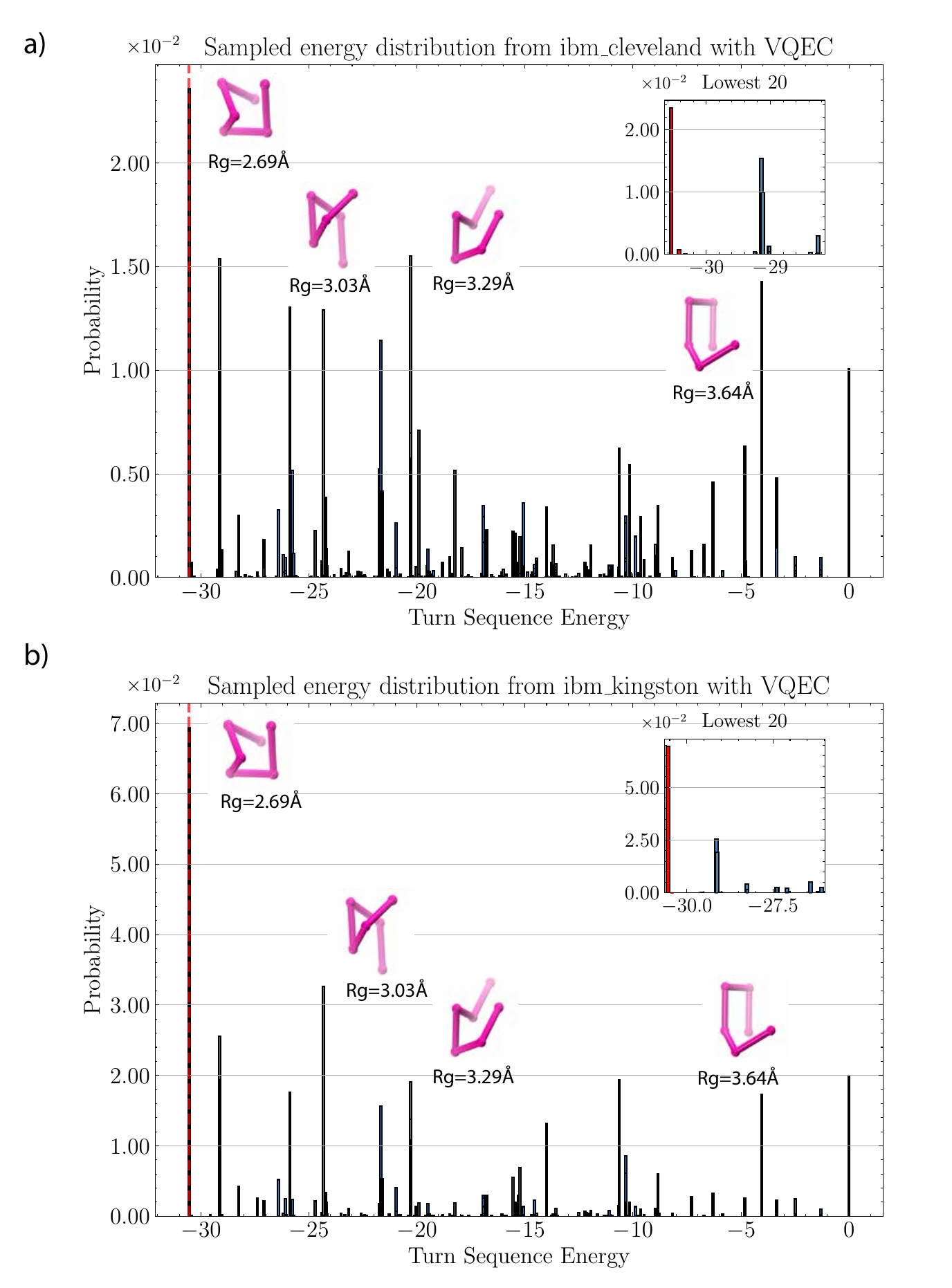}
    \caption{\small Turn sequence probabilities for \texttt{KLVFFA} ensembles when using VQEC on a) \texttt{ibm\_cleveland} and b) \texttt{ibm\_kingston}. The red dashed line indicate the ground state energy found by the classical exhaustive search. In both machines, we observe at least one of the ground state turn sequence bitstrings having the highest probability within the distribution, while presenting other clusters of high probability, higher energy bitstrings throughout the distribution. When compared to the results from the PolyFit approach, VQEC shows a more robust sampling, leading to a more concentrated distribution towards the best solution.}
    \label{vqec-plot}
\end{figure}

To diversify and benchmark our hardware experiments on different generations of quantum processors, we executed our jobs on \texttt{ibm\_cleveland} (using a 127 qubit Eagle R3 processor) and \texttt{ibm\_kingston} (using a 156 qubit Heron R2 processor), comparing the performances. We observe a significant increase in the sampling quality on the newer generation Heron processor, that is discussed in further details in the rest of this section.
We primarily use quantum computers to efficiently sample from the ground state solution to the problem Hamiltonian, whereas first, we run a training routine to optimize the quantum circuit parameters on \texttt{Qiskit-aer} high-performance simulator as described in Sec. \ref{sec:sim-results} in further details. 
The dual experiment provides an opportunity to compare optimization and sampling performance, as well as the effect of noise in terms of recovering the best possible solutions.  
Both versions of the algorithm were tested, employing the PolyFit (Fig. \ref{cheb-plot}) and VQEC (Fig. \ref{vqec-plot}). To mitigate the noise, various error mitigation and suppression methods are also applied; Pauli twirling \cite{wallman2016noise} to transform the noise channel into a Pauli noise channel and dynamical decoupling (DD) \cite{ezzell2023dynamical} to reduce and limit the unwanted interactions between qubits (usually referred to as cross-talks).

Figs.~\ref{cheb-plot}a and b illustrate the probabilities for the individual turn sequence bitstrings appearing in the sampled solutions.  
The bitstrings are sorted by their energies on the x-axis, with the lowest energy turn sequences (ground states) being on the left and higher energy sequences towards the right.  
As with the original Miyazawa-Jernigan paper~\cite{Miyazawa1996}, these energies are in units of RT (the ideal gas constant R multiplied by the temperature T), with each of these units corresponding to approximately 0.6 kcal/mol.  
For the experiment on \texttt{ibm\_cleveland}, we observe that the highest probability turn sequence bitstring is in fact the ground state solution, so the sampling process recovers the best turn sequence successfully. Moreover, despite low probabilities, we also observe the recovery of the top nine turn sequences out of the top ten. 
The same is also observed with the experiment on \texttt{ibm\_kingston}, however the ground state is found with almost twice as much probability relative to \texttt{ibm\_cleveland} (0.0175 versus 0.01).  
The other obvious trend that is observed is in the clustering of the non-ground state solutions.  
For both QPUs, we observe similar clusters in energy between -5 and -10, -11 and -20, and -22 and -30.
While there are subtle differences within these clusters, the location of most predominant peaks seem to coincide between the two plots.  
Aside from these intermediate clusters, we observe that the four largest peaks are also consistent in both cases, meaning that both QPUs recover the same four predominant turn sequence bitstrings.  For each of these, an alpha carbon PDB file was generated using the respective XYZ file produced from the hardware results.  Images of these structures are placed near each respective peak, along with a measure of their radius of gyration ($R_{g}$).
This metric is commonly used in structural biology and biophysics, serving as an indication of how open (or elongated) versus how closed (or compact) a protein structure is.  
In other words, the lower this value, the more folded the protein is (for more details on this metric and how it is calculated, please see appendix \ref{app:rg}).  
With this in mind, a very interesting trend can be observed.  
First, we can see the gradual increase in probabilities from right to left, in these 4 bitstrings.  
This means that the optimization is in fact moving in the right direction: \textit{probabilities are increasing as energy is decreasing}.  
Even more interesting is the observed decrease in $R_{g}$ in this same direction.  
These plots indicate that as probabilities increase, more compact protein structures are being produced.  
In other words, what could be represented here is not only the ground state protein structure with highest probability, but a probabilistic trajectory, \textit{modeling one of the possible folding pathways}.  
A similar trend is observed in the work of Pamidimukkala et al.~\cite{Pamidimukkala2024} (specifically, Figure 9 of their article), where $R_{g}$ is decreasing as energy is minimized throughout the VQE optimization.  
In the case of the \texttt{KLVFFA} peptide modeled here, although the known structure appears to show a more linear conformation (as seen in Fig.~\ref{fig:rmsds_workflow}b), it is also known to adopt helical and coiled conformations depending on the environment \cite{Mager2002}, some of which could be structurally similar to the more compact structures predicted with high probability in these results.  
These insightful trends could benefit from further research going forward. 

\begin{figure}[htbp]
    \centering
    \includegraphics[width=\textwidth]{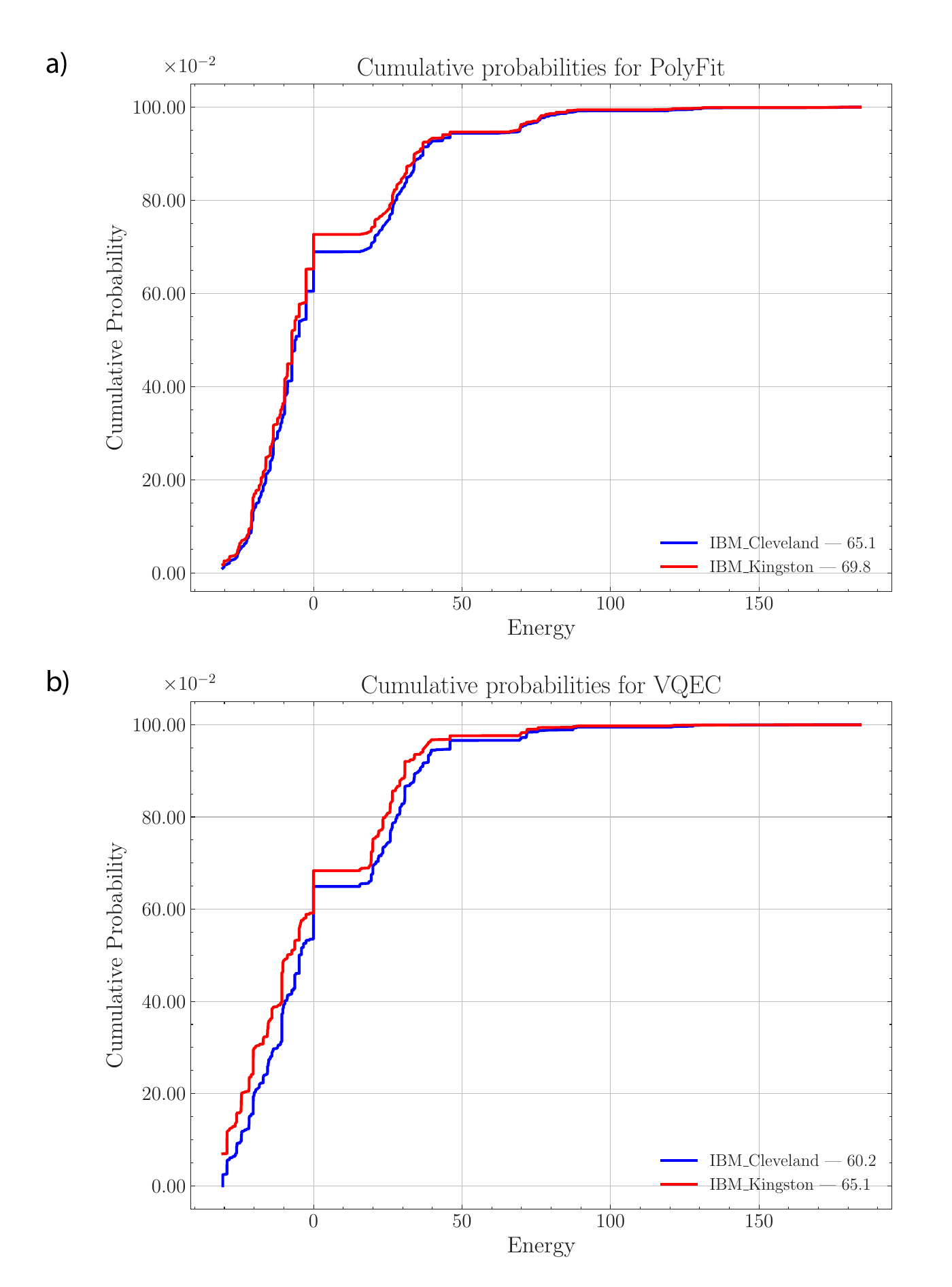}
    \caption{\small Cumulative probability plots for a) PolyFit and b) VQEC methods on \texttt{ibm\_cleveland} (blue) and \texttt{ibm\_kingston} (red). In all cases, we observe a significant percentage of the probability is accumulated in lower energies, highlighting the success of sampling.}
    \label{probs-plot}
\end{figure}

In Fig. \ref{probs-plot}a, we observe the cumulative probabilities as a function of energy. 
This plot indicates that on average, \texttt{ibm\_kingston} finds lower energy bitstrings (less than 0) in its sampled solutions, with greater probability than \texttt{ibm\_cleveland}.  
The trend is observed even through the high energy range, with a convergence of both QPUs occurring around 50 RT units in energy.  

The other technique employed in the algorithm uses VQEC \cite{leVariationalQuantumEigensolver2024}, which converts finding the global minimum problem into a saddle point detection problem. 
As mentioned previously, the main resulting difference between using either approach boils down to the scaling of the number of terms in the Hamiltonian. 
With VQEC, the number of terms remains more reasonable, but hyperparameter tuning can be challenging and the optimization is sensitive to the small changes in these parameters. 
However, the number of terms in the problem Hamiltonian for this 6 amino acid case is 2199 for VQEC construction and 18133 for the PolyFit method. 
Despite the discrepancy in the number of Hamiltonian terms, both methods are able to recover the ground state solutions successfully with relatively high probability.  
Since both methods are designed such that no slack variables are used, the total qubit number (24 qubits) does not change for PolyFit and VQEC methods implemented.

Fig. \ref{vqec-plot} summarizes the results of the quantum hardware experiments using the VQEC method. 
At first glance, an immediate difference compared to the previous results (Fig. \ref{cheb-plot}) is the significantly higher probability of finding one of the ground state solutions when compared to using the PolyFit approach.  
Specifically, for the experiment on \texttt{ibm\_cleveland} (Fig. \ref{vqec-plot}a), the highest probable ground state increases by a factor of almost 2.5x (from approximately 0.01 to 0.024 or so, in the largest red peak). 
Similarly, we see that for the experiment on \texttt{ibm\_kingston} (Fig. \ref{vqec-plot}b), there is a 4x increase in this probability (from roughly 0.0175 to 0.07).
There is a noticeable difference in the clustering of suboptimal bitstrings between both QPUs, in particular the cluster located within -20 and -32 in energy.  
For one, we see an inverted relationship between the structures presenting an $R_{g}$ of 3.03 and 3.28 \AA, respectively (the second and third structures from left to right).  
In \texttt{ibm\_cleveland}, the bitstring corresponding to the third structure has a higher probability than the second.  
This indicates that a higher energy bitstring, corresponding to a more unfolded peptide, is preferentially sampled in this case.  The relative high probabilities in these suboptimal clusters could be due to sampling noise.
The opposite can be seen on \texttt{ibm\_kingston}, where there is more than a 50\% increase in the lower energy bitstring (corresponding to the second, more folded structure, with a $R_{g}$ of 3.03 \AA). 
Thus, when comparing the overall trend of the probabilities of all four of these bitstrings, a more gradual increase from right to left (higher energy to lower energy) is observed with \texttt{ibm\_kingston}, indicating a higher quality sampling.  

In Figure \ref{probs-plot}b, we see a similar trend where the probabilities accumulate faster on \texttt{ibm\_kingston} in lower energies for VQEC. 
The difference in the cumulative probabilities between both QPUs is much more pronounced with VQEC compared to Polyfit, reaffirming what was observed in the previous plots.  
These results collectively dictate that \texttt{ibm\_kingston} is outperforming \texttt{ibm\_cleveland} with respect to optimization and sampling near the ground state.

The data in these plots highlights a few key findings.  
For one, there is a significant advantage in using VQEC when it comes to sampling the ground state turn sequences.  
For both QPUs there was a 2 to 4 fold increase in these probabilities, compared to using the PolyFit method.  
While the same number of qubits would be required, the benefit of this improved sampling, coupled with the reduced overhead with less Hamiltonian terms and the improved results, makes VQEC an attractive approach to this problem.  
In either case, \texttt{ibm\_kingston} proved to yield the best results, in terms of both the magnitudes of the highest probability states as well as the skewness of the distribution itself towards lower energy states.  
As mentioned earlier, these experiments involved using dynamical decoupling (DD) as part of the error suppression protocol to ensure the best outcomes.  
It is worth mentioning that experiments were also run without DD.  
We observed that the results improved significantly for \texttt{ibm\_cleveland} when DD was used, while there was negligible improvement for \texttt{ibm\_kingston}, regardless of the method employed (PolyFit versus VQEC).
This is somewhat expected due to improvements in Heron R2 chip which make it less susceptible to cross-talks.

\section{Discussion}
The nascent field of QPSP has been dominated by approaches based on coarse-grained lattice models, as a means to balance the trade-off between computational efficiency and the ability to accurately represent protein structures.
Therefore, within the coarse-grained framework, the choice of lattice model plays a crucial role in the success of predicting realistic protein structures.
From our analysis, it is clear that the FCC lattice yields more accurate structures than the tetrahedral lattice, especially in the presence of secondary structures.
In particular, our initial fitting analyses demonstrated that $\alpha$-helices are challenging to model with a tetrahedral lattice, given the high pitch between turns. 
Of the 11 peptides that were modeled, ranging from 10 to 26 residues in length, FCC consistently yielded lower RMSDs for these models, further proving the benefit from the added structural flexibility in allowing 12 turns per lattice node (versus 4 in the tetrahedral lattice). 

In the meantime, the increased granularity of the FCC lattice also presents more challenges in terms of the number of qubits required to encode the problem Hamiltonian, as well as the difficulty in constructing the Hamiltonian itself.
In particular, a key challenge in the development of the Hamiltonian was the need to enforce the non-overlapping constraint $H_\text{olap}$, which traditionally is handled with slack variables, leading to the requirement of extra ancilla qubits.
In this work, we utilized two methods to circumvent the need for extra qubits, namely the PolyFit and VQEC methods.
The PolyFit method works by approximating the non-overlapping constraint, which corresponds to a discrete penalty functional $F[D_{mn}]$, where $D_{mn}$ is the squared distance between the $m$-th and $n$-th residues, with a polynomial function.
This polynomial function can then be easily mapped to Pauli operators, which is then incorporated into the problem Hamiltonian.
However, the downside of this method is that it leads to an exponential increase in the number of terms in the Hamiltonian, as the degree of the polynomial increases.
Therefore, while the PolyFit method is able to handle the non-overlapping constraint without requiring extra qubits, it comes at the cost of a large number of Hamiltonian terms, which limits the scalability of the approach.
On the other hand, the VQEC method uses the Lagrangian duality to reformulate the problem as a saddle point detection problem, without the need to explicitly construct penalty functionals for the non-overlapping constraints.
Instead, the VQEC method incorporates such inequality constraints in the Lagrangian operator, which is then optimized by minimizing it with respect to the primal variables (circuit parameters) and maximizing with respect to the dual variables (Lagrange multipliers).
This allows us to avoid the exponential growth in the number of Hamiltonian terms, making the VQEC method more scalable.
However, the VQEC method requires careful tuning of the hyperparameters, such as the perturbation and update step sizes, as well as a better understanding of the cost landscape, to ensure a better convergence to the saddle point.

We have successfully implemented both methods and tested them on quantum computers for a small peptide sequence (\texttt{KLVFFA}).
After running the training routines on a noiseless simulator, we executed the optimized circuits on two different quantum processors, \texttt{ibm\_cleveland} and \texttt{ibm\_kingston}, to sample from the ground state solutions.
The results from the hardware experiments show that both methods are able to recover the ground state solutions with high probability, with VQEC outperforming PolyFit in terms of sampling quality.
Even under the noise of the quantum hardware, we observed that VQEC is able to recover the ground state solutions with a significantly higher probability than PolyFit, with a 2- to 4-fold increase in the probabilities of the ground state turn sequences.
This highlights the potential of VQEC as a more scalable and efficient method for solving the QPSP problem on quantum computers, especially for larger peptides where the number of Hamiltonian terms can become prohibitively large with the PolyFit method.
However, it is worth noting that the VQEC method requires careful tuning of the hyperparameters in the optimization process, which can be challenging and time-consuming.  
This is an area that requires further research to develop more efficient optimization strategies.
Another observation worth mentioning is that improved sampling probabilities of the ground state solution was achieved when using \texttt{ibm\_kingston} over \texttt{ibm\_cleveland}, by a factor of 2 to 3.
This highlights the advancements in quantum hardware technology, as the newer generation Heron R2 processor is able to provide better sampling quality and lower noise levels compared to the older Eagle R3 processor.

In all cases, we observed a gradual folding of the peptide structure as probability increased, and energy decreased.  
This in itself was further proof that the Hamiltonian encoding was implemented correctly and the scoring function that goes into the interaction Hamiltonian was serving its purpose.  
The optimization is guided by a minimization of energy, which in the context of scoring with contact energies such as those from the Miyazawa-Jernigan energy matrix, this simultaneously means maximization of nearest neighbor contacts.  
The combination of contacts that leads to the lowest energy yields the optimal structures.  
Thus, the optimization trajectory in the turn sequence probabilities observed on quantum hardware, could be representative of one of the possible probabilistic folding pathways of the \texttt{KLVFFA} peptide, further validating the developed method.  

In summary, we presented the first step towards an efficient quantum algorithm for protein structure prediction using the face-centered cubic (FCC) lattice, which is capable of producing more accurate C$\alpha$ lattice models than the previously studied tetrahedral and cubic lattices.
We introduced two methods to handle the non-overlapping constraints in the FCC Hamiltonian, which allow us to circumvent the need for ancilla qubits due to slack variables.
We showed the success of both methods in recovering the ground state solutions for a small peptide sequence on quantum computers, with VQEC outperforming PolyFit in terms of sampling quality.
Going forward, method-wise, we believe that the VQEC method presents a promising avenue for solving the coarse-grained QPSP problem on quantum computers due to its scalability, but further research is needed to have a better theoretical understanding of the approach in order to develop more efficient optimization algorithms.
From a broader perspective, how to find a balance between the complexity of the lattice model and the efficiency of the quantum algorithm remains a key challenge in the field of QPSP.
If quantum computing will play a role in this field, the gap between efficient energy minimization methods and more realistic modeling of proteins in the underlying problem Hamiltonian needs to be addressed, with a greater emphasis on the latter.
We believe that the FCC lattice is a key step in this direction and we hope this work will allow researchers to further explore protein structure prediction with quantum computers.


 \section{Funding}
Work on “Protein Conformation Prediction with Quantum Computing” is supported by Wellcome Leap as part of the Quantum for Bio (Q4Bio) Program.

\newpage

\begin{appendices}

\section{Prior studies on QPSP} \label{app:prior_work}
Below in Table \ref{lattice_models}, we summarize the prior work on coarse-grained lattice models for QPSP to date.

\begin{table}[htbp]
    \centering
    \footnotesize
    \renewcommand{\arraystretch}{1.2}
    \begin{tabularx}{\textwidth}{|c|c|X|}
        \hline
        \textbf{Work} & \textbf{Lattice model} & \textbf{Further details} \\
        \hline
        Perdomo \textit{et al.}~(2008) \cite{num_1} & 2D grid & {\raggedright Coordinate-based encoding on the $N \times N$ grid on a $D$-dimensional lattice, $D(N-2)\log N$-many qubits to encode the entire sequence} \\
        \hline
        Perdomo-Ortiz \textit{et al.}~(2012) \cite{num_2} & 2D grid & {\raggedright Turn-based encoding on 2D lattice, 2-qubits per turn} \\
        \hline
        Babbush \textit{et al.}~(2014) \cite{num_3} & 2D grid & {\raggedright Turn-based encoding on 2D lattice, $2N-5$ bits needed for turn encoding, drops the $\log N$ factor of spatial mapping from \cite{num_1}} \\
        \hline
        Babej \textit{et al.}~(2018) \cite{num_4} & 3D cubic lattice & {\raggedright The first 3D turn-based encoding implemented, 6 directions each requiring 3 qubits. $3N-8$ total for turn qubits} \\
        \hline
        Fingerhuth \textit{et al.}~(2018) \cite{num_5} & 3D cubic lattice & {\raggedright One hot encoded, cubic lattice using 6 qubits to represent each turn, $6N-17$ total qubits for encoding turns} \\
        \hline
        Mulligan \textit{et al.}~(2019) \cite{num_6} & N/A & {\raggedright Focuses on protein design, but presents an off-lattice side-chain packing model for protein folding} \\
        \hline
        Robert \textit{et al.}~(2021) \cite{num_7} & 3D tetrahedral lattice & {\raggedright First paper to introduce tetrahedral model, $2(N-3)$ with dense encoding for configuration qubits, total qubit scaling $\mathcal{O}(N^2)$ on alternating sublattices} \\
        \hline
        Wong \& Chang~(2021) \cite{num_8} & 3D BCC lattice & {\raggedright Body-centered cubic (BCC) lattice with the HP model, Grover's search algorithm proposed to find the lowest energy} \\
        \hline
        Outeiral \textit{et al.}~(2021) \cite{num_9} & N/A & {\raggedright A meta-study to understand the limitations of annealing for QPSP} \\
        \hline
        Wong \& Chang~(2022) \cite{num_10} & 2D grid & {\raggedright Grover's algorithm on a simulator implemented, $\mathcal{O}(N^3)$ scaling for qubits} \\
        \hline
        Irb\"ack \textit{et al.}~(2022) \cite{irback2022folding} & 2D square lattice & {\raggedright A novel coordinate-based encoding with the HP model; the resulting QUBO Hamiltonian can be solved on quantum annealers} \\
        \hline
        Kahatami \textit{et al.}~(2023) \cite{num_11} & 2D square lattice & {\raggedright Grover-based search algorithm proposed} \\
        \hline
        Chandarana \textit{et al.}~(2023) \cite{num_12} & 3D tetrahedral lattice from \cite{num_7} & {\raggedright Implements a digitized counterdiabatic approach for solving QPSP} \\
        \hline
        Boulebnane \textit{et al.}~(2023) \cite{num_13} & 3D tetrahedral lattice from \cite{num_7} & {\raggedright Encoding non-backtracking into the problem rather than the Hamiltonian for better scaling} \\
        \hline
        Linn \textit{et al.}~(2023) \cite{num_14} & N/A & {\raggedright A meta-study of various lattice and off-lattice side-chain packing models with different encoding such as unary, binary, and BU-binary} \\
        \hline
        Doga \textit{et al.}~(2024) \cite{doga2024perspective} & 3D tetrahedral lattice from \cite{num_7} & {\raggedright Computational hardness analysis for QPSP problem based on CASP results and classical state-of-the-art methods} \\
        \hline
        Romero \textit{et al.}~(2025) \cite{romero2025proteinfoldingalltoalltrappedion} & 3D tetrahedral lattice from \cite{num_7} & {\raggedright Implemented using the BF-DCQO algorithm, using all-to-all connectivity on a trapped-ion system} \\
        \hline
    \end{tabularx}
    \caption{\small A summary of coarse-grained lattice models used in the literature.}
    \label{lattice_models}
\end{table}

\section{FCC encoding details}\label{app:encoding_details}

Below we show the mapping of the 12 distinct turns on the FCC lattice to bitstrings, i.e., qubit configurations:

\begin{table}[h] 
\centering
\fontsize{10pt}{10pt}\selectfont
\begin{tabular}{ c  c  c}
    Turn direction & Encoding ($q_i q_{i+1} q_{i+2} q_{i+3}$) & Turn label \\
     \hline
     (1,1,0) & 0000 & 0\\ 
     (-1,-1,0) & 0011 & 1\\ 
     (-1,1,0) & 1100 & 2\\
     (1,-1,0) & 1111 & 3\\
     (0,1,1) & 1001 & 4\\
     (0,-1,-1) & 0101 & 5\\
     (0,1,-1) & 1010 & 6 \\
     (0,-1,1) & 0110 & 7\\
     (1,0,1) & 1000 & 8\\
     (-1,0,-1) & 0100 & 9\\
     (1,0,-1) & 1011 & a\\
     (-1,0,1) & 0111 & b\\
     \hline \\
     \hspace{1em}
\end{tabular}
\caption{Mapping of the 12 distinct turns on the FCC lattice to bitstrings and their corresponding turn labels.}
\label{tab:fcc_turns}
\end{table}

Turn indicator functions for the 12 distinct turns are defined as follows:
\begin{align*}
    d^j_{(+x,+y)} &= (1-q_{\phi+2})(1-q_{\phi+3})(1-q_{\phi+4})(1-q_{\phi+5}), \\
    d^j_{(-x,-y)} &= (1-q_{\phi+2})(1-q_{\phi+3})q_{\phi+4}q_{\phi+5}, \\
    d^j_{(-x,+y)} &= q_{\phi+2}q_{\phi+3}(1-q_{\phi+4})(1-q_{\phi+5}), \\
    d^j_{(+x,-y)} &= q_{\phi+2}q_{\phi+3}q_{\phi+4}q_{\phi+5}, \\
    d^j_{(+y,+z)} &= q_{\phi+2}(1-q_{\phi+3})(1-q_{\phi+4})q_{\phi+5}, \\
    d^j_{(-y,-z)} &= (1-q_{\phi+2})q_{\phi+3}q_{\phi+4}(1-q_{\phi+5}), \\
    d^j_{(+y,-z)} &= q_{\phi+2}(1-q_{\phi+3})q_{\phi+4}(1-q_{\phi+5}), \\
    d^j_{(-y,+z)} &= (1-q_{\phi+2})q_{\phi+3}q_{\phi+4}(1-q_{\phi+5}), \\
    d^j_{(+x,+z)} &= q_{\phi+2}(1-q_{\phi+3})(1-q_{\phi+4})(1-q_{\phi+5}), \\
    d^j_{(-x,-z)} &= (1-q_{\phi+2})q_{\phi+3}(1-q_{\phi+4})(1-q_{\phi+5}), \\
    d^j_{(+x,-z)} &= q_{\phi+2}(1-q_{\phi+3})q_{\phi+4}q_{\phi+5}, \\
    d^j_{(-x,+z)} &= (1-q_{\phi+2})q_{\phi+3}q_{\phi+4}(1-q_{\phi+5}),
\end{align*}
where $\phi = 4(j-2)$, and $j=2, \cdots, N-2$ is the index of the turn.

The turn indicator functions for the 4 unused turns are defined as follows:
\begin{align*}
    d^j_{0010} &= (1-q_{\phi+2})(1-q_{\phi+3})q_{\phi+4}(1-q_{\phi+5}), \\
    d^j_{0001} &= (1-q_{\phi+2})(1-q_{\phi+3})(1-q_{\phi+4})q_{\phi+5}, \\
    d^j_{1101} &= q_{\phi+2}q_{\phi+3}(1-q_{\phi+4})q_{\phi+5}, \\
    d^j_{1110} &= q_{\phi+2}q_{\phi+3}q_{\phi+4}(1-q_{\phi+5}).
\end{align*}

With the turn indicator functions defined, we can now compute the position of each amino acid in the protein sequence.
For example, to compute the $x$ coordinate of the amino acid at position $m$, we need to sum up the contributions from all previous turns, starting from the first one.
For each turn, we add the contributions from the turn indicator functions that correspond to the $(+x, +y)$, $(+x, -y)$, $(+x, +z)$, and $(+x, -z)$ turns, and subtract the contributions from the $(-x, +y)$, $(-x, -y)$, $(-x, +z)$, and $(-x, -z)$ turns.
We summarize the position functions for the $x$, $y$, and $z$ coordinates of the amino acid at position $m$ in the following equations:
\begin{align*}
    x_m &= 1 + (1-q_0)(1-q_1)+q_0(1-q_1)-(1-q_0)q_1-q_0q_1 \\
    &\hphantom{=\ } + \sum_{j=2}^{m-1}\qty[d^j_{(+x,+y)} + d^j_{(+x,-y)} + d^j_{(+x,+z)} + d^j_{(+x,-z)} - d^j_{(-x,+y)} - d^j_{(-x,-y)} - d^j_{(-x,+z)} - d^j_{(-x,-z)}], \\
    y_m &= 1 + (1-q_0)(1-q_1)+q_0q_1 \\
    &\hphantom{=\ } + \sum_{j=2}^{m-1}\qty[d^j_{(+x,+y)} + d^j_{(-x,+y)} + d^j_{(+y,+z)} + d^j_{(+y,-z)} - d^j_{(+x,-y)} - d^j_{(-x,-y)} - d^j_{(-y,+z)} - d^j_{(-y,-z)}], \\
    z_m &= q_0(1-q_1)-(1-q_0)q_1 \\
    &\hphantom{=\ } + \sum_{j=2}^{m-1}\qty[d^j_{(+x,+z)} + d^j_{(-x,+z)} + d^j_{(+y,+z)} + d^j_{(-y,+z)} - d^j_{(+x,-z)} - d^j_{(-x,-z)} - d^j_{(+y,-z)} - d^j_{(-y,-z)}].
\end{align*}

\section[Calculating radius of gyration Rg and root mean square deviation (RMSD)]{Calculating radius of gyration ($R_{g}$) and root mean square deviation (RMSD)}\label{app:rg}
Alpha carbon PDB files were generated from the quantum algorithm's XYZ files.  GROMACS was then used to setup the initial files for each protein structure \cite{Abraham2015}, and then for calculating their mass-weighted radius of gyration ($R_{g}$).

For any given protein, $R_{g}$ is a measure of how extended or compact the structure is, with respect to its center of mass.  Thus, in the context of PSP, it can be used to compare the degree of folding in a series of models.  The mass-weighted radius of gyration \( R_g \) is given by:

\[
R_g = \sqrt{ \frac{1}{M} \sum_{i=1}^{N} m_i \left\| \mathbf{r}_i - \mathbf{r}_{\text{com}} \right\|^2 }
\]

where:
\begin{itemize}
  \item \( m_i \) is the mass of atom \( i \),
  \item \( \mathbf{r}_i \) is the position vector of atom \( i \),
  \item \( \mathbf{r}_{\text{com}} = \frac{1}{M} \sum_{i=1}^{N} m_i \mathbf{r}_i \) is the center of mass,
  \item \( M = \sum_{i=1}^{N} m_i \) is the total mass.
\end{itemize}
\hfill
\break

To directly compare the models to some sort of reference structure, such as the optimal lattice model or experimental structure, RMSDs can be calculated following an alignment of these structures.
In structural bioinformatics, this metric helps quantify the similarity between two molecular structures.  The lower this value, the more accurate the model.  Given two structures with $N$ equivalent atoms, where $\mathbf{x}_i$ and $\mathbf{y}_i$ denote the Cartesian coordinates of the $i$-th atom in the reference and target structure respectively, the RMSD is defined as:
\[
\text{RMSD} = \sqrt{\frac{1}{N} \sum_{i=1}^{N} \left\| \mathbf{x}_i - \mathbf{y}_i \right\|^2}
\]
We use modules available in Biopython \cite{Cock2009} to carry out these calculations, but there are several bioinformatics tools that can be leveraged for this task.

\section{Details on the exhaustive search protocol}\label{app:ex_search}
In this section we discuss the implementation of the exhaustive search protocols for both the Tetrahedral and FCC algorithms.  In general, the algorithms are written in C++, performing a classical exhaustive search over all valid 3D self-avoiding walks of a short peptide on each lattice, scoring them using a pairwise amino acid contact energy model (e.g., the Miyazawa–Jernigan matrix).  In either case, the objective is to find the lowest energy 3D conformation of any given peptide sequence. For each lattice, a description of the exhaustive search algorithm protocol is described below.

\subsection{Tetrahedral lattice}
\begin{enumerate}

    \item \textbf{Read Inputs:} Defines peptide sequence length via \texttt{\#define AMINOACIDS} and its string via \texttt{\#define PEPTIDE}, and load the energy matrix.

    \item \textbf{Map Amino Acids to Energy Indices:} For each amino acid in the sequence, determine its corresponding index in the energy matrix.

    \item \textbf{Initialize Tetrahedral Lattice Geometry and Turn Sequence Vectors}
      \begin{itemize}
        \item Lattice composed of tetrahedral vectors:
        \[
        \left( \pm1, \pm1, \pm1 \right)/\sqrt{3}
        \]
        \item Alternates between sublattice A (odd indices) and B (even indices).
        \item Encodes directionality and step parity to determine spatial coordinates.
      \end{itemize}
    
    \item \textbf{Transform Turn Sequence Representation}
      \begin{itemize}
        \item Uses 3-turn relative directions $\{0, 1, 2\}$ to define local conformations \cite{num_13}.
        \item Converts relative turns to absolute turns using:
        \[
        \texttt{turnseq\_abs}[i] = (\texttt{turnseq\_rel}[i] + \texttt{turnseq\_abs}[\texttt{aa\_parent}[i]] + 1) \bmod 4
        \]
        \item Encodes directions as bit-shifted chunks in 64-bit integers to sppedup coordinate accumulation.
      \end{itemize}
    
    \item \textbf{Enumerate all Permutations and Search Over Configuration Space:}
      \begin{itemize}
        \item Generates all possible 3-turn sequences of length $N-3$.
        \item First turn is fixed.
        \item The number of permutations is $3^{N-3}$, with N being the number of amino acids.
        \item Looping over every permutations, using OpenMP to parallelize over turn sequence permutations.
      \end{itemize}
    
    \item \textbf{Check for Collisions and Compute Nearest-Neighbor Interaction Energies:} For each valid configuration, compute the objective functions below:
        Let the peptide consist of $N$ amino acids, indexed by $i, j \in \{1, \ldots, N\}$. The total energy of a folding conformation is computed by summing the pairwise interaction energies $E_{ij}(d_{ij})$ over non-adjacent residue pairs:

       \[
        E_{\text{total\_1NN}} = 
        \sum_{\substack{0 \leq i < j < N \\ j - i \geq 5 \\ (j - i) \bmod 2 = 1}} 
        \delta_{d_{ij}^2, 1} \cdot C_{ij} \cdot \varepsilon_{ij}
        \]
          \[
        E_{\text{total\_2NN}} = 
        \sum_{\substack{0 \leq i < j < N \\ j - i \geq 5}} 
        \left[
            \delta_{d_{ij}^2, 1} \cdot C_{ij}^{(1)} \cdot \varepsilon_{ij}^{(1)}
            + \delta_{d_{ij}^2, 2} \cdot C_{ij}^{(2)} \cdot \varepsilon_{ij}^{(2)}
        \right]
        \]
      \begin{itemize}
        \item For all potentially interacting pairs $(i, j)$ with $|i - j| \geq 4$, calculate accumulated lattice vectors.  \(d_{ij}^2\) squared lattice distance between residues \(i\) and \(j\), calculated from the cumulative lattice walk:
        \[
        d_{ij}^2 = n_0^2 + n_1^2 + n_2^2 + n_3^2
        \]
        where \(n_0, n_1, n_2, n_3\) are lattice direction components in the 4D representation of the tetrahedral lattice. 
        
        \item If $d_{ij}^2 = 0$, a collision or overlap between the amino acids is found.  If $d_{ij}^2 = 1$, the pair forms a 1st nearest neighbor (1NN) interaction.  If $d_{ij}^2 = 2$ the pair forms a 2nd nearest neighbor (2NN). 
        
        \item \(\varepsilon_{ij}^{(1)}\), \(\varepsilon_{ij}^{(2)}\): pairwise interaction energies from the energy matrix (e.g. Miyazawa-Jernigan contact energies), applied for 1NNs and 2NNs.  The latter (and beyond) is scaled according to lattice neighbor level and chosen lattice.  For case of overlaps, a large penalty is applied to the \texttt{obj} argument.  These relationships are summarized below.

            \[
        E_{ij}(d_{ij}) =
        \begin{cases}
        \infty & \text{if } d_{ij}^2 = 0 \quad (\text{collision}) \\
        \varepsilon_{ij} & \text{if } d_{ij}^2 = 1 \quad (\text{1st nearest neighbor}) \\
        \lambda \cdot \varepsilon_{ij} & \text{if } d_{ij}^2 = 2 \quad (\text{2nd nearest neighbor})
        \end{cases}
        \]
        \[
        \varepsilon_{ij}^{(2)} = \lambda \cdot \varepsilon_{ij}^{(1)}, \quad \text{with } \lambda = {\sqrt{3/8}} \quad \text{on the tetrahedral lattice}
        \]

        \item \(\delta_{d_{ij}^2, 1}\), \(\delta_{d_{ij}^2, 2}\): Kronecker delta functions indicating 1NN and 2NN conditions, activating either scenario, respectively, as seen below.
        \[
        \delta_{d_{ij}^2, d^2} =
        \begin{cases}
            1, & \text{if } d_{ij}^2 = d^2 \\
            0, & \text{otherwise}
        \end{cases}
        \quad \text{for } d^2 \in \{1, 2\}
        \]
    
        \item \(C_{ij}^{(1)}\), \(C_{ij}^{(2)}\): Collision filters applied for 1NN and 2NN, respectively.
        \[
        C_{ij}^{(k)} =
        \begin{cases}
            1, & \text{if no coordinate collision with neighbors of residue } j \\
            0, & \text{if collision occurs}
        \end{cases}
        \quad \text{for } k \in \{1, 2\}
        \]
    
        \item The constraint \(j - i \geq 5\): Prevents counting local contacts, consistent with self-avoiding walk and physical constraints of the peptide chain.  This is applied in both the 1NN scenario and beyond.

        \item The constraint $(j - i) \bmod 2 = 1$ is applied only to the 1NN case.  This is due to the tetrahedral lattice being a bipartite lattice, and 1NN interactions consequently only permitted between beads on alternating sublattices (based on the inequality previously mentioned inequality above).  For 2NN interactions, the residues can either be on the same or different sublattices.  Thus, the equation for $E_{total\_2NN}$ above, takes on a slighly different form, where the summation does not check for this constraint and this check is bypassed in the exhaustive search algorithm.
        
        \item Configurations with collisions are further rejected by comparing forward and backward lattice state (i.e., configurations with two consecutive, equivalent sublattice turns).
      \end{itemize}

    \item \textbf{Tracking Best Conformations}
      \begin{itemize}
        \item The energies are computed on the fly, with a continuous update of the top-$K$ solutions (corresponding to the lowest $K$ energy conformations), while discarding the rest in an effort to save on memory usage.  \item Each thread maintains its best and top-$K$ energy conformations. After parallel execution, global best and top-$K$ structures are compiled and sorted.
      \end{itemize}
    
    \item \textbf{Output Generation}
      \begin{itemize}
      \item Two major outputs are generated: a \texttt{topobj.txt} file storing all saved conformers and associated meta data, and XYZ files for each conformer.
        \item \texttt{topobj.txt} contains:
          \begin{itemize}
            \item conformer energy
            \item absolute turn sequence
            \item qubit bitstring configuration
            \item XYZ coordinates for each conformer
          \end{itemize}
        \item 3D coordinates for each conformation are saved in \texttt{.xyz} format, scaled by 3.8~\text{\AA} in between lattice nodes, to reflect the known average distance between adjacent alpha carbons in nature.
    
      \end{itemize}

    \end{enumerate}

\subsection{FCC lattice}

    \begin{enumerate}
        \item \textbf{Read Input:} Load the hardcoded amino acid sequence length and string (for example, \#define AMINOACIDS 6 and \#define PEPTIDE "KLVFFA"), and the energy matrix (e.g., from the CSV file of the Miyazawa-Jernigan matrix).
        
        \item \textbf{Map Amino Acids to Energy Indices:} For each amino acid in the sequence, determine its corresponding index in the energy matrix.
        
        \item \textbf{Initialize FCC Steps:} Define all possible step directions on the FCC lattice using a lookup table of 12 directions and their inverses, encoded as 3D vectors (specified in the \texttt{step\_lookup[]} argument).  Importantly, the following restrictions are enforced:
        \begin{enumerate}
            \item First turn: fixed. 
            \item Second turn: only 4 unique directions (due to rotational symmetry). 
            \item Subsequent turns: 11 options (excluding strict backtracking)
        \end{enumerate}
        
        \item \textbf{Enumerate All Permutations:} Systematically generate all valid non-backtracking permutations of lattice directions (up to $ 4 \times 11^{N-3}$, where $N$ is the number of amino acids). Each next step has 11 valid directions and is generated such that it is not in the opposite direction as the prior step.  The \texttt{turnseq\_abs[]} array encodes each of these specific walks on the lattice (the sequence of turn directions from the \texttt{step\_lookup[]} table).
        
        \item \textbf{Generate Lattice Path:} For each permutation, calculate the stepwise 3D coordinates of the peptide on the FCC lattice.

        \item \textbf{Check for Collisions and Compute Nearest-Neighbor Interaction Energies:} For each valid configuration, compute the objective functions below:
        Let the peptide consist of $N$ amino acids, indexed by $i, j \in \{1, \ldots, N\}$. The total energy of a folding conformation is computed by summing the pairwise interaction energies $E_{ij}(d_{ij})$ over non-adjacent residue pairs:
    
    \[
    E_{\text{total}} = \sum_{\substack{i < j \\ j - i \geq 2}} E_{ij}(d_{ij})
    \]
    
    where $d_{ij}$ is the squared distance on the FCC lattice between residues $i$ and $j$.  Depending on the nearest neighbor interaction in question, $E_{ij}(d_{ij})$ can take one of several forms, as defined below:
    
    \[
    E_{ij}(d_{ij}) =
    \begin{cases}
    \infty & \text{if } d_{ij}^2 = 0 \quad (\text{collision}) \\
    \varepsilon_{ij} & \text{if } d_{ij}^2 = 2 \quad (\text{1st nearest neighbor}) \\
    \frac{1}{\sqrt{2}} \varepsilon_{ij} & \text{if } d_{ij}^2 = 4 \quad (\text{2nd nearest neighbor})
    \end{cases}
    \]

        \[
    \varepsilon_{ij}^{(2)} = \lambda \cdot \varepsilon_{ij}^{(1)}, \quad \text{with } \lambda = \frac{1}{\sqrt{2}} \quad \text{on the FCC lattice}
    \]
    
    Here:
    \begin{itemize}
        \item $\varepsilon_{ij}$ is the interaction energy between amino acids $i$ and $j$ from the chosen energy matrix.
        \item $d_{ij}^2 = (x_j - x_i)^2 + (y_j - y_i)^2 + (z_j - z_i)^2$ is the squared lattice distance between the two residues.
        \item For interactions greater than 1NN, a scaling factor is applied which is designed to mimic the natural decay of the interaction energy in an inverse distance dependent manner.  This relationship is weighted relative to the 1NN interaction, which leads to a scaling factor of $\frac{1}{\sqrt{2}}$ in the case of 2NN interactions (where $\sqrt{2}$ corresponds to the 1NN euclidian distance on the lattice).  In this sense, the strength of the interaction is being ``tapered off'' off accordingly as the distance on the lattice increases.
        \item If a collision is detected ($d_{ij}^2 = 0$), this results in a large, user-defined penalty being imposed (depicted as infinitely positive above, but in reality specified as a real number such as 10,000) to help discard unrealistic configurations containing overlaps.  This is set with the \texttt{obj} argument in the algorithm.
    \end{itemize}
    
    The final energy of the conformation is then:
    
    \[
    E_{\text{total}} = \sum_{\substack{i < j \\ j - i \geq 2}} \left[ \delta_{d_{ij}^2, 2} \cdot \varepsilon_{ij} + \delta_{d_{ij}^2, 4} \cdot \frac{1}{\sqrt{2}} \varepsilon_{ij} \right]
    \]
    
    where $\delta_{d_{ij}^2, d_{NN}}$ is the Kronecker delta, which serves as an activating function that only adds the interaction energy if the respective nearest neighbor distance for residue pairs \textit{i} and \textit{j} is satisfied.
    
    \[
    \delta_{d_{ij}^2, d_{NN}} =
    \begin{cases}
    1 & \text{if } d_{ij}^2 = d_{NN} \\
    0 & \text{otherwise}
    \end{cases}
    \]
    
    The current exhaustive search algorithms employed in our work can account up to 2NN interactions, but the method is readily extendable to longer range interactions as well. In any case, what changes is the magnitude of the scaling factor as the nearest neighbor distance increases, as seen below.
    
    \begin{table}[h!]
    \centering
    \begin{tabular}{|c|c|c|c|}
    \hline
    \textbf{Interaction Level} & \textbf{Squared Distance \( d^2 \)} & \textbf{Euclidean Distance \( d \)} & \textbf{Scaling Factor} \\
    \hline
    1st NN & 2 & \( \sqrt{2} \) & \( 1 \) \\
    \hline
    2nd NN & 4 & \( 2 \) & \( \frac{1}{\sqrt{2}} \approx 0.7071 \) \\
    \hline
    3rd NN & 6 & \( \sqrt{6} \) & \( \frac{1}{\sqrt{3}} \approx 0.577 \) \\
    \hline
    \end{tabular}
    \caption{Scaling factors for nearest neighbor interactions on an FCC lattice.}
    \end{table}

    \item \textbf{Update Best and Top Configurations:} As with the tetrahedral exhaustive search algorithm, the objective is maintaining the lowest-energy (top-$K$) conformers across parallel threads.  To save memory, the conformational energies are calculated on the fly, and only the Top-$K$ configurations are saved, with this list continuously being updated as lower energy conformers are produced.
    
    \item \textbf{Output Generation}
      \begin{itemize}
      \item Two major outputs are generated: a \texttt{topobj.txt} file storing all saved conformers and associated meta data, and XYZ files for each conformer.
        \item \texttt{topobj.txt} contains:
          \begin{itemize}
            \item conformer energy
            \item unique turn sequence alphanumeric string
            \item XYZ coordinates for each conformer
          \end{itemize}
    \item 3D coordinates for each conformation are saved in \texttt{.xyz} format, scaled by 3.8~\text{\AA} in between lattice nodes, to reflect the known average distance between adjacent alpha carbons in nature.
    \end{itemize}
\end{enumerate}

\begin{figure}[t]
    \centering
    \includegraphics[scale=0.35]{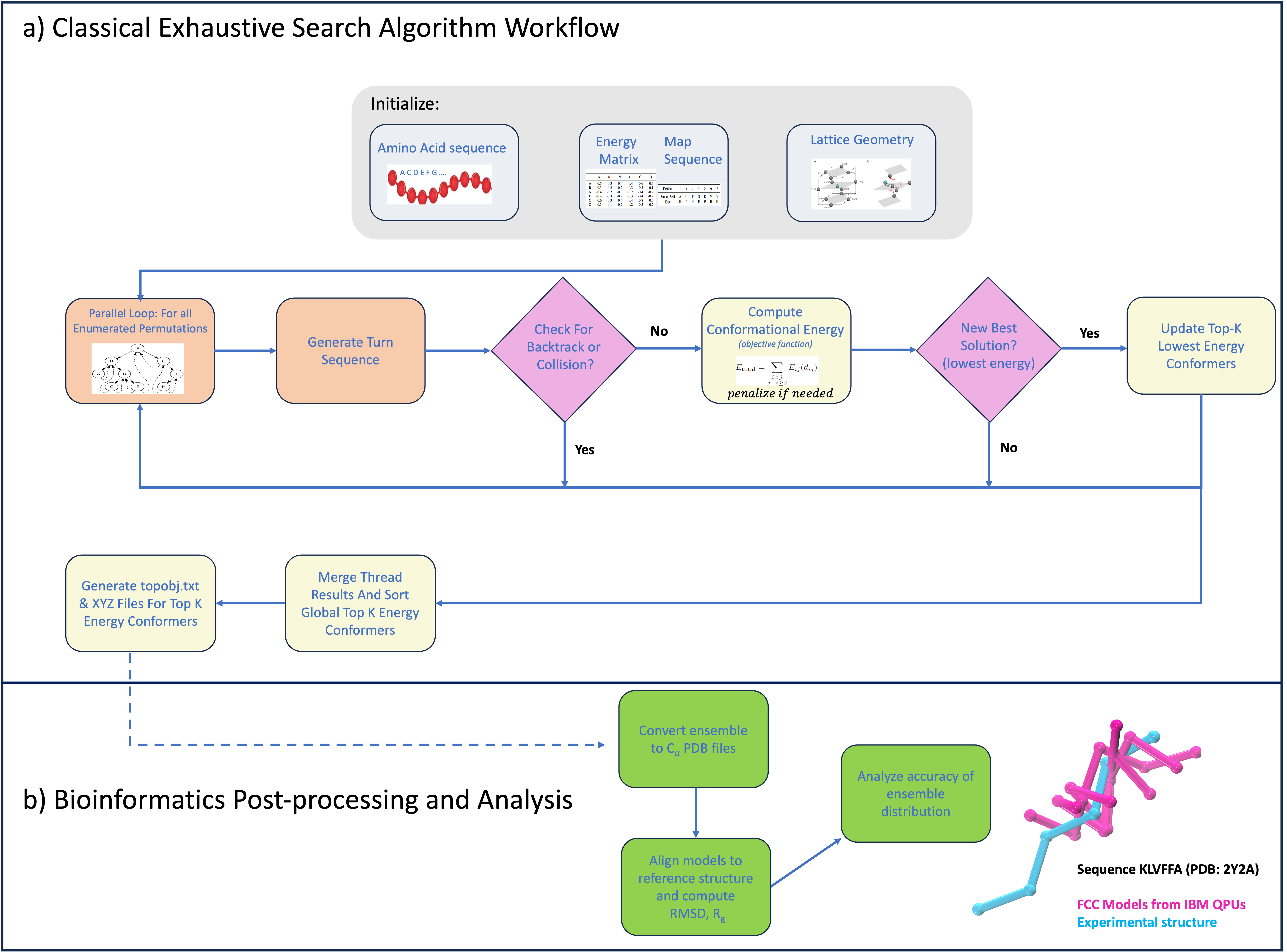}
    \caption{\small a) Schematic of the classical search algorithm workflow and  b) a fundamental set of postprocessing steps in bioinformatics analysis.  At their core, both the tetrahedral and FCC search algorithms follow the workflow in a).  The steps in b) do not include all-atom reconstruction and refinements with molecular mechanics force fields (publishing these longer, more sophisticated workflows is part of our ongoing work to be published in subsequent studies).  These post-processing steps are equally compatible with the XYZ files from either one of the classical exhaustive search algorithms or the actual quantum algorithms (ours and the original tetrahedral code from ~\cite{robertResourceefficientQuantumAlgorithm2021}), as long as the coordinates are scaled to bond lengths of 3.8 \r{A} in between each alpha carbon (by default in our FCC quantum algorithm presented here, and the exhaustive search algorithms).  Additional images inserted in some each of the workflow steps were reproduced from \cite{miyazawa1985estimation, lau1989lattice}}
    \label{fig:rmsds_workflow}
\end{figure}

\end{appendices}

\newpage 

\bibliographystyle{unsrtnat}
\bibliography{bibliography}

\end{document}